\documentclass[aps,floats,prl,
twocolumn]{revtex4-1}

\usepackage{amsfonts,amsmath} \usepackage{bm} \usepackage{dcolumn}
\usepackage{epsfig} \usepackage{latexsym}

\providecommand{\gtrless}{
  \mathrel{% it's a relation
  \smash{% we don't want that it influences the interline spacing
  \vcenter{% the symbol will be vertically centered
    \offinterlineskip % no interline skip here
    \ialign{% build a table
       \hfil##\hfil\cr % just one centered column
       $>$\cr % first row
       \noalign{\kern-.3ex}% shorten the vertical distance
       $<$\cr % second row
    }% end of the \ialign
  }% end of \vcenter
  }% end of \smash
  \vphantom{>}% pretend it's as high as a >
  }% end of \mathrel
}

\usepackage{soul}
\usepackage[dvipsnames]{xcolor}

\begin{document}

\title{Invariance principle for wave propagation inside inhomogeneously disordered materials}

\author{Yiming Huang,$^{1\dagger}$ Chushun Tian,$^{2\dagger}$ Victor A. Gopar,$^{3\ast}$ Ping Fang,$^4$ and Azriel Z. Genack$^{1\ast}$}

\affiliation{$^1$Department of Physics, Queens College and The Graduate Center of the City University of New York, Flushing, NY 11367\\
$^2$CAS Key Laboratory of Theoretical Physics and Institute
of Theoretical Physics, Chinese Academy of Sciences, Beijing 100190, China\\
$^3$Departamento de F\'{\i}sica Te\'{o}rica and BIFI, Universidad de Zaragoza, Pedro Cerbuna 12, 50009 Zaragoza, Spain\\
$^4$School of Science, Beijing University of Posts and Telecommunications, Beijing 100876, China}

\begin{abstract}

Disorder is more the rule than the exception in natural and synthetic materials. Nonetheless, wave propagation within inhomogeneously disordered materials has received scant attention. We combine microwave experiments and theory to find the spatial variation of generic wave propagation quantities in inhomogeneously disordered materials. We demonstrate {that wave statistics within samples of any dimension are independent of the detailed structure of a material and depend only on the net strengths of distributed scattering and reflection between the observation point and each of the boundaries.}

\end{abstract}

\date{\today}

\maketitle

Coherent waves launched into a random sample are scrambled in a haze of scattered fields that is manifested in the speckle pattern of scattered light. This makes disordered materials opaque and impairs communication, imaging, and transport. Studies of transport in random systems have focused on disordered materials with uniform scattering strength throughout the medium and possible surface reflection \cite{Sheng06,Akkermans07,Rotter17,Rossum99,Mello-book,Anderson79,Dorokhov82,Mello88}. For example, Brownian motion of particles in a homogeneous liquid was shown by Einstein to be due to randomly fluctuating forces on microparticles by molecules in thermal equilibrium \cite{Einstein56}, while in Chandrahekar's radiative transfer model, the rate of particle scattering out of or into a particular direction is homogeneous \cite{Chandrasekhar}. In quantum scattering, the suppression of electron diffusion was first studied in the Anderson tight-binding model, in which the distributions of electron energy at each lattice site are the same \cite{Anderson58}. Localization of radiowaves was first found in calculations for random waveguides with a spatially homogeneous distribution of disorder in the dielectric constant \cite{Gertsenshtein59}. Such studies have led to a deep understanding of ballistic and diffusive propagation and of Anderson localization. However, the paradigm of homogeneously disordered materials does not represent key elements of our surroundings, which are generally inhomogeneously disordered with a spatially varying scattering strength and multiple interfaces within the bulk of the material and at the surface. Understanding wave propagation in such materials would enable a wide range of applications in diverse fields, including medical imaging \cite{Gagnon08}, lidar and remote sensing \cite{Alfano17}, astrophysics \cite{Chandrasekhar}, telecommunication \cite{Sklar97}, electronics \cite{Nazarov94,Bliokh15}, phononics \cite{Balandin11,Li12,Lepri03}, invisibility cloaking \cite{Wegner16}, and photothermal therapy \cite{photothermal}.

In this letter, we go beyond canonical homogeneously disordered materials. We begin by considering a simple departure from the traditional uniformly disordered material -- the insertion of a single reflector into a uniformly disordered sample. We discover an invariance principle for such materials: Wave statistics at any the point of observation is invariant with regard to displacement of a reflector along the length of the sample, apart from a discontinuity that arises when the reflector crosses that point. This leads to the discovery of an even broader invariance principle. By that principle we obtain a full description of wave propagation and statistics for general inhomogeneously disordered materials with multiple embedded reflectors {or tunneling barriers and a longitudinally varying transport mean free path $\ell(x)$}.

\begin{figure*}[t]
\includegraphics[width=2\columnwidth] {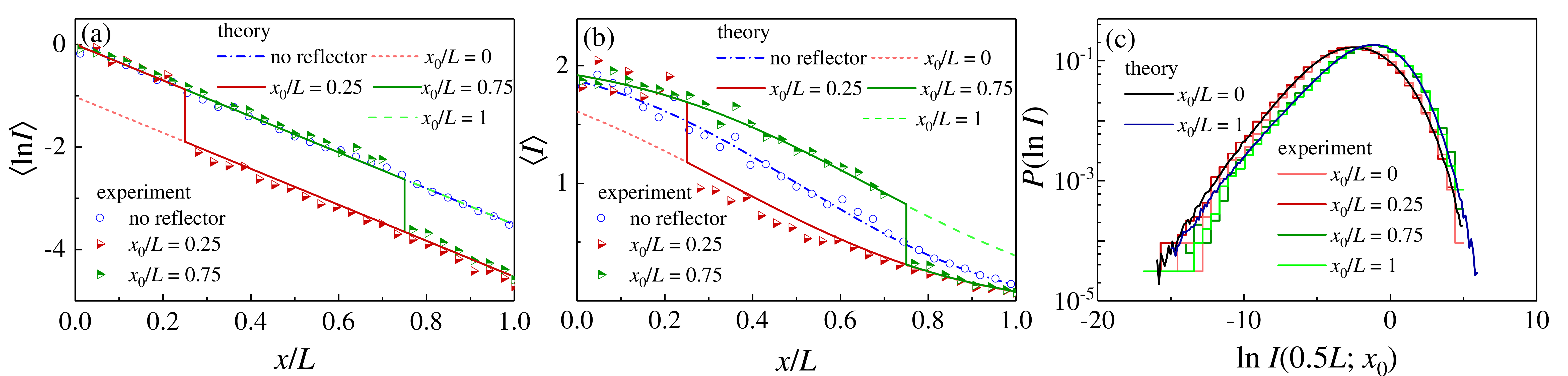}
\caption{Measurements of and theoretical results for $O$$($$x$$;$$x_0$$)$$=$$\langle$$\ln I$$($$x$$;$$x_0$$)$$\rangle$ (a), $\langle$$I$$($$x$$;$$x_0$$)$$\rangle$ (b) and $P$$($$\ln I$$($$0.5L$$;$$x_0$$)$$)$ (c) in {$1$D} samples with a reflector at different $x_0$ or without a reflector are in good agreement. They all exhibit the invariance and discontinuity with regard to displacement of the reflector, and are identical to $O$$($$x$$;$$L$$($$0$$)$$)$ for $x$$<$$x_0$ ($x$$>$$x_0$).}
\label{fig:2}
\end{figure*}

We carry out microwave measurements in a uniformly disordered one-dimensional ($1$D) open medium of length $L$ with a reflector at depth $x_0$.
%(Fig.~\ref{fig:1}).
We consider the intensity at a point $x$, $I$$($$x$$;x_0$$)$, which is normalized so that its value at the output is equal to the flux transmission coefficient, for an ensemble of disorder configurations. The results of measurements of the ensemble averages $\langle$$\ln I$$($$x$$;$$x_0$$)$$\rangle$, $\langle$$I$$($$x$$;$$x_0$$)$$\rangle$ and the probability density function $P$$($$\ln I$$($$x$$;$$x_0$$)$$)$ are explained using random matrix theory (RMT) \cite{Mello15,Genack17,Mello-book} and supersymmetry field theory (SUSY) \cite{Tian08,Tian10}. We consider $P$$($$\ln I$$)$ rather than $P$$($$I$$)$ since $\ln I$ self-averages and so a comparison to theory can be made with measurements on fewer disorder configurations. RMT and SUSY address wave propagation from different perspectives and together provide a universal description for it.

We find that the average and the probability density function of all local quantities that can be expressed in terms of the Green function of the wave equation share a common feature with regard to their variation with $x_0$: When $x$ is fixed, but $x_0$ is changed, a quantity $O$$($$x$$;$$x_0$$)$ representative of waves in a disordered ensemble is unchanged as long as $x$ remains in front of or behind the reflector, and changes discontinuously from $O$$($$x$$;$$0$$)$ to $O$$($$x$$;$$L$$)$ when $x_0$ crosses $x$. This behavior is encapsulated in the following identity proved below:
\begin{eqnarray}\label{eq:1}
    O(x;x_0)=\theta(x_0-x)O(x;L)+\theta(x-x_0)O(x;0),
\end{eqnarray}
where $\theta$ is the Heaviside function. This allows us to find the spatial structure of {the statistics of intensity and of the intensity in transmission eigenchannels \cite{Mosk08,Choi12,Mosk12,Genack15,Cao16}}. The impact of distributed scattering and discrete reflection on the profile of $\langle$$\ln I$$\rangle$ is additive, while the profile of $\langle I\rangle$ is determined by the position-dependent diffusion coefficient $D$$($$x$$;$$x_0$$)$, which is a function of the strength of distributed scattering, discrete reflection, and the relative positions of $x$ and $x_0$. $D$$($$x$$;$$x_0$$)$ not only satisfies Eq.~(\ref{eq:1}), but also obeys a scaling law,
\begin{equation}\label{eq:4}
    D(x;x_0)/D_0=D_\infty(\lambda(x;x_0)),
\end{equation}
once it is rescaled by the Boltzmann diffusion coefficient $D_0$. Here $x$ and all parameters describing the inhomogeneity enter into the scaling factor $\lambda(x;x_0)$, and the scaling function $D_\infty(\lambda)$ describes propagation in both homogeneously and inhomogeneously disordered materials. These findings go beyond traditional studies that address the interplay between surface reflection and particle scattering, which ignore wave interference \cite{Rossum99,Chandrasekhar,Lagendijk89,Zhu91,Genack93,Lagendijk97}. {The results are surprising since the field at $x$ results from the coherent superposition of waves arriving at $x$ which is inevitably affected by the specific location of a reflector, $x_0$, being in front of or behind $x$. Finally, the principle applies not only in $1$D, but also in quasi-$1$D and higher dimension.}

Measurements are carried out with the use of a vector network analyzer in a single-mode rectangular copper waveguide containing ceramic slabs of thickness $6.6$ mm and Teflon U-channel air spacers with thickness randomly chosen from three values: $1.27$, $2.55$ and $3.82$ cm. The sample of length $L$$=$$86.0$ cm is contained in a copper waveguide with a cutoff frequency of $6.56$ GHz. Successive sample elements are selected randomly with a probability of
$1/2$ for the ceramic slabs and $1/6$ for each thickness of the U-channel elements. Spectra are taken for ensembles of $100$ random
configurations without a reflector and with a reflector at relative depths of $x_0$$/$$L$$=$$0$$,$$1/4$$,$$3/4$$,$$1$. The reflector is a thin copper plate covering $76\%$ of the waveguide cross section with transmission coefficient $\Gamma$$=$$0.36$ in the empty waveguide.

The wave is detected by an antenna inserted sequentially into a series of holes of diameter ${3.17}$ mm spaced by ${1}$ cm along the waveguide.
%(Fig.~\ref{fig:1}).
Field spectra are taken over the frequency interval $10.00$-${10.70}$ GHz over which the waveguide supports a single mode. The incident
intensity $I_{0}$ is found by
fitting the expression for the intensity of counter-propagating waves:
%$I(x)$$=$
$I_{0}$$[$$1$$+$$r^2$$+$$2$$r$$\cos$$($$2$$k$$x$$+$$\varphi$$)$$]$ in a $4$-cm-long segment before the random sample to determine the incident field.
Here $r$ is the magnitude of the reflected
field, $k$ is the wavenumber determined from waveguide dispersion, and $\varphi$ is the phase. We undo the impact of absorption using the method described in the Supplementary Materials (SM) \cite{SM}. Measurements of intensity at each position are normalized by measurements in the empty waveguide with absorbers at the ends of the waveguide so that the sensitivity is uniform at all positions.

Measured profiles of $\langle\ln I(x;x_0)\rangle$ for samples without a reflector and with a reflector at $x_0$$=$$L/4$ or $3L/4$, in which the impact of absorption is removed are plotted in Fig.~\ref{fig:2}a. $\langle$$ \ln I$$\rangle$
decreases linearly in front of the
reflector and overlaps the profile
measured in samples without a reflector, but drops sharply behind the
reflector and then continues to fall with the same slope as before the reflector. The profiles in samples with a reflector at $x_0$$=$$L/4$ and $3L/4$ are identical, in the region in front of ($x$$<$$L/4$) and behind ($x$$>$$3L/4$) the reflectors, demonstrating that the size of the drop is independent of $x_0$. Away from $x_0$, the slope of $\langle$$ \ln I$$\rangle$ in all regions is the same as the slope in a sample without a reflector.
When $x$ is rescaled by $L$, this gives a slope of magnitude $s$$=$$L/\ell$$=$$3.51$ \cite{Genack17} corresponding to a mean free path $\ell$$=$$24.5$ cm. Below and in the SM, we use RMT to show
\begin{eqnarray}
\label{averlnIz}
\langle \ln I(x;x_0) \rangle = \left\{
 \begin{array}{ll}
  -x/\ell  & \textrm{for $x < x_0$  }\\
   -x/\ell +\ln \Gamma & \textrm{for $x > x_0$}
 \end{array}. \right.
\end{eqnarray}{The discontinuity at $x_0$ is equal to the logarithm of the transmission coefficient $\Gamma$ of the reflector.}
This analytic result is plotted in Fig.~\ref{fig:2}a, and seen to be in excellent agreement with measurements. Equation (\ref{averlnIz}) is a special case of the general result of Eq.~(\ref{eq:1}). For $x$$<$$x_0$ ($x$$>$$x_0$), $\langle$$\ln I$$($$x$$;$$x_0$$)$$\rangle$ is the same as when the reflector is at the output (input).

\begin{figure}[b]
\includegraphics[width=\columnwidth] {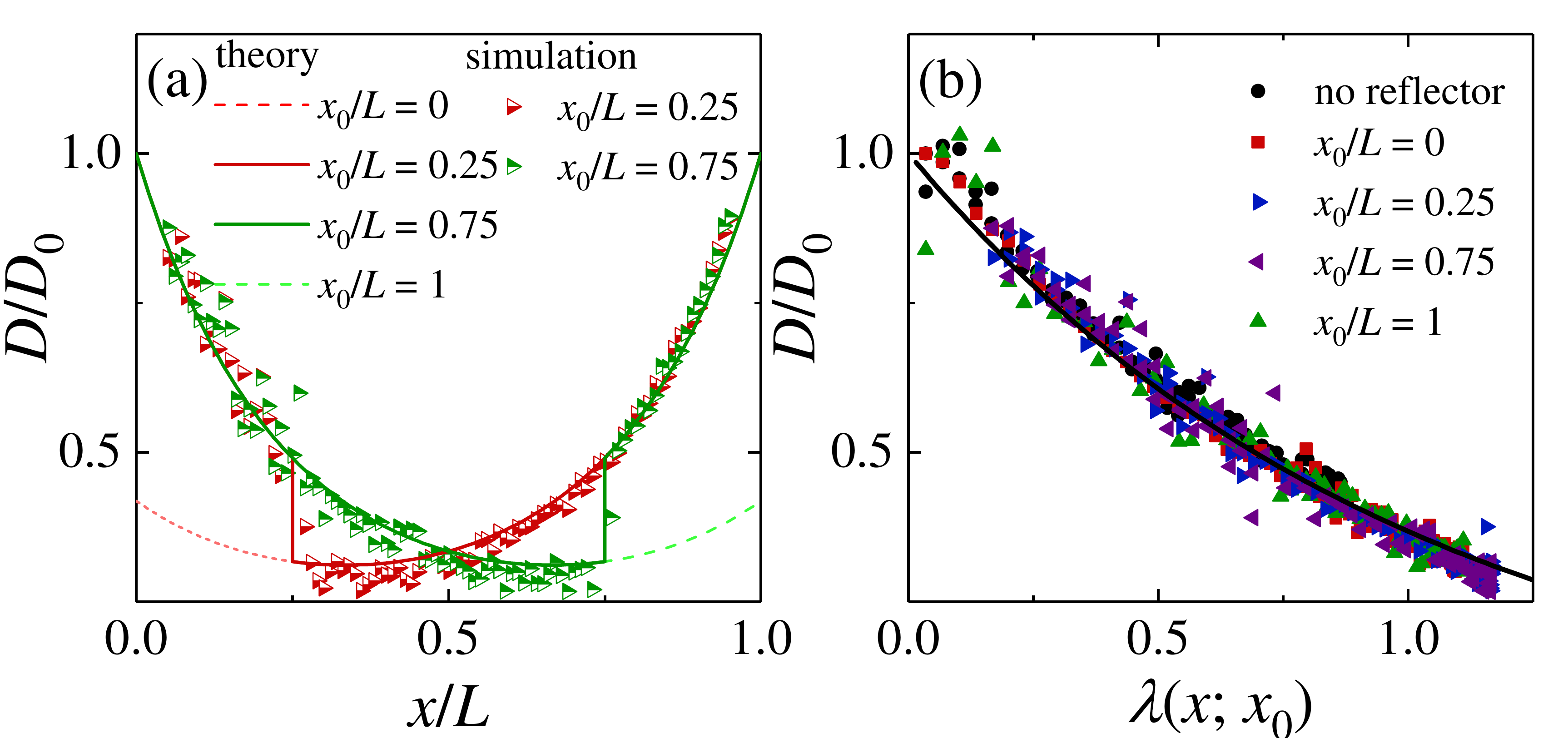}
\caption{(a) Simulation and analytic results for {$1$D} samples with a reflector at different $x_0$ show that the profiles $D$$($$x$$;$$x_0$$)$ overlap $D$$($$x$$;$$L$$($$0$$)$$)$ for $x$$<$$x_0$ ($x$$>$$x_0$) and display a discontinuity at $x_0$. (b) Upon mapping $x$ to $\lambda$$($$x$$;$$x_0$$)$, simulated profiles of $D$$($$x$$;$$x_0$$)$$/$$D_0$ (symbols) for samples with different $x_0$ or without a reflector collapse to a single curve: $D_\infty$$($$\lambda$$)$$=$$e^{-\lambda}$ (solid line).}
\label{fig:3}
\end{figure}

Measured profiles of $\langle I(x;x_0)\rangle$ for samples without a reflector and with a reflector at $x_0$$=$$L/4$ or $3L/4$ are plotted in Fig.~\ref{fig:2}b. The profiles for the reflector at $x_0$$=$$L/4$ and $3L/4$ still overlap for $x$$<$$L/4$ and $x$$>$$3L/4$, as is the case for $\langle$$\ln I$$\rangle$. Though $\langle$$ \ln I$$($$x$$;$$x_0$$)$$\rangle$ in front of the reflectors ($x$$<$$L/4$) is not affected by the presence of the reflector, $\langle$$I$$($$x$$;$$x_0$$)$$\rangle$ increases significantly for $x$$<$$L/4$ relative to the profile for samples without a reflector. Using SUSY we show that as an another application of Eq.~(\ref{eq:1}),
\begin{equation}\label{eq:28}
\langle I(x;x_0)\rangle=\theta(x_0-x)\langle I(x;L)\rangle+\theta(x-x_0)\langle I(x;0)\rangle,
\end{equation}
where
\begin{eqnarray}\label{eq:12}
    \langle I(x;0(L))\rangle
    =\langle I(0;0(L))\rangle-\left(\langle I(0;0)\rangle+\langle I(0;L)\rangle-2\right)\nonumber \\
   \times\frac{{\rm erf}\left(\sqrt{\frac{sL}{L+\zeta}}\frac{2x-(L\mp \zeta)}{2L}\right)
   -{\rm erf}\left(-\sqrt{\frac{sL}{L+\zeta}}\frac{L\mp \zeta}{2L}\right)}
   {{\rm erf}\left(\sqrt{\frac{sL}{L+\zeta}}\frac{L\pm \zeta}{2L}\right)-{\rm erf}\left(-\sqrt{\frac{sL}{L+\zeta}}\frac{L\mp \zeta}{2L}\right)}.\quad
\end{eqnarray}
Here $\zeta$ is a length determined by $\Gamma$ and $\langle$$I$$($$0$$;$$0$$)$$\rangle$$\equiv$$\langle$$I$$($$0^+$$;$$0$$)$$\rangle$. From Eq.~(\ref{eq:12}), we find $\langle$$I$$($$x$$;$$0$$)$$\rangle$$+$$\langle$$I$$($$L$$-$$x$$;$$L$$)$$\rangle$$=$$2$. This is in accordance with the sum of intensity at a point due to all incident channels being equal to the local density of states with its average unaffected by the presence of a reflector. Good agreement of Eqs.~(\ref{eq:28}) and (\ref{eq:12}) with measurements is found for $s$$=$$3.51$, $\zeta$$=$$0.33L$ \cite{SM}, $\langle$$I$$($$0$$;$$0$$)$$\rangle$$=$$1.61$ and $\langle$$I$$($$0$$;$$L$$)$$\rangle$$=$$1.92$, with the last two parameters calculated using Eq.~(\ref{eq:13}) below. For samples without a reflector, $\zeta$ vanishes. In this case, we find with the use of Eq.~(\ref{eq:13}) that $\langle$$I$$($$0$$;$$0$$)$$\rangle$$=$$\langle$$I$$($$0$$;$$L$$)$$\rangle$$=$$1.86$. Substituting this into Eq.~(\ref{eq:12}) gives $\langle$$I$$($$x$$;$$0$$)$$\rangle$$=$$\langle$$I$$($$x$$;$$L$$)$$\rangle$ with a profile in good agreement with measurements.

Measurements of $P$$($$\ln I($$L$$/$$2$$;$$x_0$$)$$)$ at $x_0$$/$$L$$=$$0$,$1$$/$$4$, $3$$/$$4$, $1$ are shown in Fig.~\ref{fig:2}c and compared to RMT calculations for $P$$($$\ln I$$($$L$$/$$2$$;$$0(L)$$)$$)$. We see that the measurements of $P$$($$\ln I$$)$ for $x_0$$/$$L$$=$$0$$,$$1$$/$$4$ coincide with the theoretical
result of $P$$($$\ln I$$($$L$$/$$2$$;$$0$$)$$)$, while, for $x_0$$/$$L$$=$$1$$,$$3$$/$$4$, they match the theoretical
result for $P$$($$\ln I$$($$L$$/$$2$$;$$L$$)$$)$. This is in accordance with Eq.~(\ref{eq:1}), which predicts that $P$$($$\ln I$$($$L$$/$$2$$;$$x_0$$)$$)$ is invariant with regard to the displacement of the reflector as long as $\theta$$($$x$$-$$x_0$$)$ does not change, but changes suddenly when $x_0$ crosses $L$$/$$2$.

To study universal aspects of $\langle$$I$$\rangle$, we show below that
\begin{equation}\label{eq:3}
    -\partial_xD(x;x_0)\partial_x\langle I\rangle=0
\end{equation}
in the interior of the medium. This differs from the normal diffusion equation in the position dependence of the diffusion coefficient, which is the result of the spatial variation of localization effects in open media \cite{Tiggelen00,Tian08,Skipetrov08,Tian10,Yamilov14}.
%We stress that $D$$($$x$$;$$x_0$$)$ is an intrinsic quantity of the system in the spirit of linear response theory.
%This is unlike other dependences upon position \cite{Zeldovich66}
%arising from the dependence of the diffusion coefficient on the solution of the generalized diffusion equation.
Below we will show that
\begin{eqnarray}
\label{eq:2}
\frac{D(x;x_0)}{D_0}\!=\!e^{-\lambda(x;x_0)},
\lambda(x;x_0)\!=\!\left\{\begin{array}{c}
                      \frac{(x+\zeta)(L-x)}{\ell (L+\zeta)},x>x_0  \\
                      \frac{x(L+\zeta-x)}{\ell (L+\zeta)},x<x_0
                    \end{array},\,\,\,\right.
\end{eqnarray}
where $\lambda$ is essentially the probability density of return obtained from the diffusion equation with diffusion coefficient $D_0$. Because
$\lambda$ for $x$$>$$x_0$ ($x$$<$$x_0$) is the same as when $x_0$$=$$0$ ($L$), it obeys Eq.~(\ref{eq:1}), and so does $D$.

\begin{figure*}[t]
\includegraphics[width=2\columnwidth] {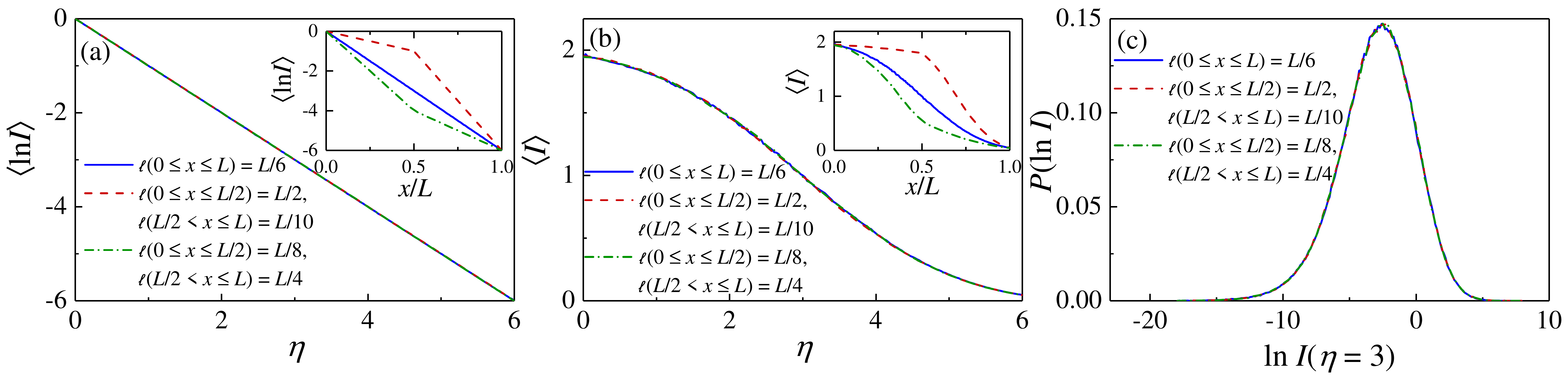}
\caption{Simulations for three ensembles, two consisting of {$1$D} samples composed of two segments of different scattering strength (red dashed and green dash-dotted curves) and one of uniformly disordered samples (blue solid curve). In all samples, $\eta_L$$=$$6$. Upon mapping $x$ to $\eta$$($$x$$)$, the profiles of $\langle$$\ln I$$\rangle$ (a, inset), $\langle$$I$$\rangle$ (b, inset), and $P$$($$\ln I$$)$ (c) in different ensembles collapse to a single curve (main panel).}
\label{fig:4}
\end{figure*}

Intensity fluctuations are too large to allow for an accurate experimental determination of $D$$($$x$$;$$x_0$$)$ with data from $100$ configurations. Thus we perform simulations using the
experimental values for $\ell$, $L$ and $\Gamma$, and compute $D$$($$x$$;$$x_0$$)$ from
$-$$\langle$$T$$\rangle$$/$$\partial_x$$\langle$$I$$\rangle$, which is the
generalized Fick's law. Here $\langle$$T$$\rangle$ is the
ensemble-averaged transmission coefficient, which is equal to the
flux. Simulation results for $x_0$$=$$L/4$ and
$3L/4$ are shown in Fig.~\ref{fig:3}a and are in
good agreement with the analytic result of
Eq.~(\ref{eq:2}). We next use the expression for $\lambda$ given
above to map $x$ to $\lambda$$($$x$$;$$x_0$$)$. For samples without a
reflector, we set $\zeta$$=$$0$ so that
$\lambda$$=$$x$$($$L$$-$$x$$)$$/$$($$\ell$$L$$)$. The simulated
profile $D$$($$x$$;$$x_0$$)$$/$$D_0$ for samples with
different $x_0$ or for samples without a
reflector collapse to a single curve:
$D_\infty$$($$\lambda$$)$$=$$e^{-\lambda}$, as shown in
Fig.~\ref{fig:3}b. This scaling law was found previously for uniformly disordered materials \cite{Tian10}.

{We sketch the analytic derivations. The full theory is provided in a self-contained manner in the SM. We first consider quasi-$1$D but locally two-dimensional ($2$D)} random media and then extrapolate the final results from quasi-$1$D to $1$D. The dielectric constant $\epsilon$$($$\boldsymbol{r}$$)$ exhibits independent Gaussian fluctuations around unity at every point $\boldsymbol{r}$$\equiv$$($$x$$,$$y$$)$. The reflector is modeled by a layer of high dielectric constant. The propagation of a wave of angular frequency $\omega$ is described by the retarded (advanced) Green function $G^{R(A)}$$\equiv$$($$\nabla^2$$+$$\omega^2$$\epsilon$$($$\boldsymbol{r}$$)$$\pm$$i$$0^+$$)^{-1}$. {Using the SUSY technique \cite{Efetov97}, $G^{R(A)}$ can be expressed in terms of a path integral over a supervector field.} Since a local quantity such as $I$$($$x$$;$$x_0$$)$ can be expressed in terms of $G^{R(A)}$, upon performing the disorder average, one can express $O$$($$x$$;$$x_0$$)$ in terms of a functional integral over a supermatrix field $Q(x)$ \cite{Tian08,Tian10}. {Upon rescaling $x$ by $4\xi$ where $\xi$$=$$\pi$$\nu$$D_0$ with $\nu$ the density of states per unit length, we find
\begin{eqnarray}
    O(x;x_0)=\int DQ(x)M_O[Q(x)]\qquad\qquad\qquad\qquad\qquad\label{eq:48}\\
    \times\left\{\begin{array}{c}
                    W(Q_L,Q(x);\tilde L-x){\tilde W}(Q(x),\Lambda;x,x_0),x>x_0\\
                    {\tilde W}(Q_L,Q(x);\tilde L-x,x_0-x)W(Q(x),\Lambda;x),x<x_0
                  \end{array}
    \right.
    \nonumber
\end{eqnarray}
with $\tilde L$$=$$L$$/$$($$4$$\xi$$)$. Different observables $O$ differ only in the functional $M_O$$[$$Q$$($$x$$)$$]$ and in the boundary constraint $Q$$($$L$$)$$=$$Q_L$ [$Q$$($$0$$)$ is a constant matrix $\Lambda$.]. Here $W$ is the heat kernel
%$W$$($$Q$$,$$Q'$$;$$\tilde L$$)$$
%\equiv$$\int_{Q(0)=Q'}^{Q(\tilde L)=Q}$$D$$[$$Q$$]$$e^{-\frac{1}{8}\int_0^{\tilde L} dx{\rm str}(\partial_x Q)^2}$
and
\begin{eqnarray}\label{eq:35}
    &&{\tilde W}(Q,Q';x_1,x_2)\equiv\int DQ_-DQ_+ e^{\gamma {\rm str}(Q_-Q_+)}\nonumber\\
    &\times& W(Q,Q_+;x_1-x_2)W(Q_-,Q';x_2),\,{\rm for}\, x_1>x_2.
\end{eqnarray}
Here $-$$\gamma$${\rm str}$$($$Q_-$$Q_+$$)$ is the action accounting for wave tunneling through the
reflector, with $\gamma$ depending on $\omega$, $\Gamma$ and the cross sectional area, and str represents the supertrace. Using Eq.~(\ref{eq:48})}, we find that both $\partial_{x_0}$$ O$$($$x$$;$$x_0$$<$$x$$)$ and $\partial_{x_0}$$ O$$($$x$$;$$x_0$$>$$x$$)$ vanish \cite{SM}. This justifies Eq.~(\ref{eq:1}).

{Applying Eq.~(\ref{eq:48}) to the spatial correlator: ${\cal Y}$$($$x$$,$$x'$$)$$\equiv$$\langle$$\int\!\!\!\!\int$$dy$$dy'$$ G^A$$($$\boldsymbol{r}$$,$$\boldsymbol{r}'$$)$$G^R$$($$\boldsymbol{r}'$$,$$\boldsymbol{r}$$)$$\rangle$, we find that it is the fundamental solution of the generalized diffusion equation, $-$$\partial_x$$D$$($$x$$;$$x_0$$)$$\partial_x$${\cal Y}$$=$$\delta$$($$x$$-$$x'$$)$, from which Eq.~(\ref{eq:3}) follows.} We also find
$D_\infty$$($$\lambda$$)$$=$$1-$$\lambda$$+$${\cal O}$$($$\lambda^3$$)$ for $\lambda$$\ll$$1$ corresponding to weakly
localized waves, with $\lambda$ given by Eq.~(\ref{eq:2}) and $\zeta
%(\Gamma)
$$=$$\xi/(4\gamma)
%\frac{\sqrt{\pi} \ell}{d}\frac{\Gamma((d+1)/2)}{\Gamma(d/2)}\frac{2-\Gamma}{\Gamma}
$. This perturbative expansion of $D_\infty$ is exactly the same for samples without a reflector \cite{Tian10}, and the presence of a reflector at $x_0$ only enters into $\lambda$. The scaling behavior described by Eq.~(\ref{eq:4}) is expected to hold for all $\lambda$. Thus we can apply the results for localized samples without a reflector \cite{Tian10} to the present case. This gives $D_\infty$$\approx$$e^{-\lambda}$ for $\lambda$$\gtrsim$$1$ and $D$$($$x$$;$$x_0$$)$$=$$D_0$$e^{-\lambda(x;x_0)}$ in $1$D. Solving Eq.~(\ref{eq:3}) gives Eq.~(\ref{eq:12}).

$\langle$$\ln I$$\rangle$ and $P$$($$\ln
I$$)$ for $1$D are most readily calculated using RMT. {By using the maximum-entropy approach \cite{Mello88} we show that \cite{SM}}
%Multiple scattering processes
%on the left (right) of $x$ are described by the transfer matrices $M_{l(r)}$,
%and the reflector is modeled by a transfer matrix $M_\Gamma$ which depends on $\Gamma$.
%Let $a$ ($b$) be the amplitude of the right(left)-going wave at $x$
%(Fig.~\ref{fig:1}). Then $I$$=$$|$$a$$e^{ikx}$$+$$b$$e^{-ikx}$$|^2$,
%which can be written in terms of $M_{l(r)}$.
%In the polar representation \cite{Mello-book}, $M_{l(r)}$ are
%parametrized by $\theta_{l (r)}, \mu_{l (r)}$, and $\lambda_{l(r)}$ (see
%SM). According to the maximum-entropy approach \cite{Mello88},
%$\theta_{l (r)}, \mu_{l (r)}$ are uniformly distributed over $[0,2\pi]$,
%while the distribution of $\lambda_{l(r)}$,
%$p_{s_l(s_r)}$$($$\lambda_{l(r)}$$)$, depends on the ratio
%$s_l$$=$$x$$/$$\ell$ [$s_r$$=$$($$L-$$x$$)$$/$$\ell$], whose explicit
%form \cite{Molina10,Kleftogiannis13} is given in the SM. This gives
\begin{eqnarray}\label{eq:13}
     P\left(\ln I(x;x_0)
\right)=\int\!\!\!\!\int_0^{2\pi}\frac{d\theta_ld\theta_r}{(2\pi)^2}
\int\!\!\!\!\int_0^{2\pi}\frac{d\mu_ld\mu_r}{(2\pi)^2}\int\!\!\!\!\int_0^\infty
d\lambda_ld\lambda_r\nonumber\\
     \times p_{s_l}(\lambda_l)p_{s_r}(\lambda_r)\delta\left(\ln I(x;x_0)
- \ln I(x, \{\lambda,\theta,\mu\})\right).\qquad
\end{eqnarray}
{Here $I$$($$x$$,$$\{$$\lambda$$,$$\theta$$,$$\mu$$\}$$)$ is the
expression for intensity in the polar
representation \cite{Mello-book}. The parameters $\theta_{l (r)}, \mu_{l (r)}$ are uniformly distributed over $[0,2\pi]$, and $\lambda_{l(r)}$ is distributed according to $p_{s_l(s_r)}$$($$\lambda_{l(r)}$$)$ \cite{Molina10,Kleftogiannis13} with $s_l$$=$$x$$/$$\ell$ [$s_r$$=$$($$L-$$x$$)$$/$$\ell$]. The explicit
forms of $I$$($$x$$,$$\{$$\lambda$$,$$\theta$$,$$\mu$$\}$$)$ and $p_{s_l(s_r)}$ are given in the SM.} From Eq.~(\ref{eq:13}), one may obtain the statistics of
any function
of intensity. It gives the values
$\langle$$I$$($$0$$;$$0$$($$L$$)$$)$$\rangle$ in Eq.~(\ref{eq:12}) and
$P$$($$\ln$$I$$)$ at $x$$=$$L$$/$$2$ plotted in Fig.~\ref{fig:2}c and the
analytic expression of Eq.~(\ref{averlnIz}).

%We have so far considered uniformly disordered samples with a single embedded reflector.
In general, inhomogeneity arise from multiple segments with different scattering strength or a smoothly varying mean free path, and multiple embedded reflectors. In this case, we map $x$ to $\eta$$=$$\eta$$($$x$$)$$=$$\int_0^x$$\frac{dx'}{\ell(x')}$ [$\eta_L$$=$$\eta$$($$L$$)$].
%, where $\ell(x)$ is the local mean free path, namely, the mean free path corresponding to a uniform disordered sample with the same disorder structure as at $x$.
A generic average quantity assumes the form, $O$$($$\eta$$;$$\{$$\eta_{i}$$\}$$)$, where $\{\eta_i\}$ are the coordinates of interfaces or reflectors. Using SUSY, we find
\begin{eqnarray}\label{eq:11}
    \forall j:\partial_{\eta_{j}}O(\eta;\{\eta_{i}\})=0,\quad {\rm if\, and\, only\, if}\, \eta_{j}\neq\eta.
\end{eqnarray}
Thus $O$ is invariant with regard to arbitrary shuffling of interfaces or reflectors, as long as they do not cross $\eta$. This generalizes Eq.~(\ref{eq:1}). Indeed, Eq.~(\ref{eq:1}) has an equivalent form, namely, $\partial_{x_0}$$O$$($$x$$;$$x_0$$)$$=$$0$ if and only if $x_0$$\neq$$x$, which is a special case of Eq.~(\ref{eq:11}).

\begin{figure}[t]
\includegraphics[width=\columnwidth] {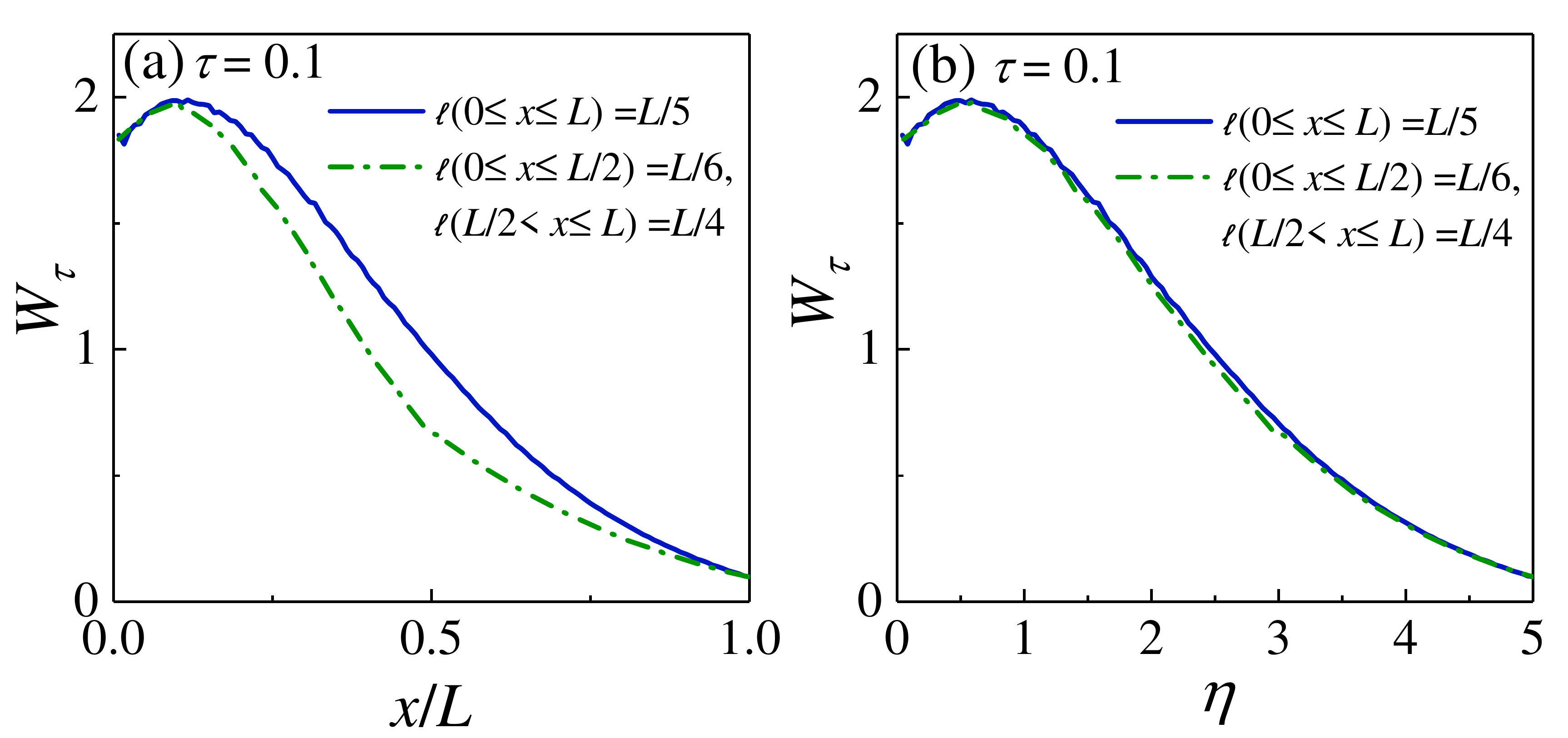}
\caption{{(a) Simulations of the eigenchannel profile $W_\tau$ for two ensembles of $2$D diffusive samples. In one ensemble, the sample is composed of two segments of different scattering strength (green dash-dotted curve) while in the other the scattering strength is uniform (blue solid curve). In all samples, $\eta_L$$=$$5$. (b) Upon mapping $x$ to $\eta$$($$x$$)$, the profiles of $W_\tau$ in the two ensembles collapse to a single curve.}}
\label{fig:5}
\end{figure}

For inhomogeneity arising solely from variations in scattering strength, Eq.~(\ref{eq:11}) implies that the $\eta$ dependence of $O$ is identical to that in a uniformly disordered sample. For $\langle$$\ln$$I$$\rangle$, we find using RMT that $\langle$$\ln$$I$$\rangle$$=$$-$$\eta$. For $\langle$$I$$\rangle$, we find using SUSY that $-$$\partial_\eta$$D$$($$\eta$$)$$\partial_\eta$$\langle$$I$$\rangle$$=$$\delta$$($$\eta$$-$$\eta'$$)$ with $D$$($$\eta$$)$$=$$e^{-\frac{\eta(\eta_L-\eta)}{\eta_L}}$, which gives
\begin{eqnarray}\label{eq:14}
    \langle I(\eta)\rangle
    &=&\langle I(0)\rangle-2\left(\langle I(0)\rangle-1\right)\nonumber \\
   &\times&\frac{{\rm erf}\left(\frac{2\eta-\eta_L}{2\sqrt{\eta_L}}\right)
   -{\rm erf}\left(-\frac{\sqrt{\eta_L}}{2}\right)}
   {{\rm erf}\left(\frac{\sqrt{\eta_L}}{2}\right)-{\rm erf}\left(-\frac{\sqrt{\eta_L}}{2}\right)}.
\end{eqnarray}
In Fig.~\ref{fig:4}, we compare numerical results for intensity statistics in samples with $\eta_L=6$ but different variations of $\ell(x)$. The spatial profiles of $\langle \ln I(\eta)\rangle$ and $\langle I(\eta) \rangle$ and the probability distributions of $\ln I$ of any value of $\eta$ are the same in different samples.

{The invariance principle for high-dimensional diffusive samples, in which nonuniformity in scattering arises either from an embedded reflector or a spatially varying mean free path, is demonstrated theoretically in the SM. In Fig.~\ref{fig:5}, we compare numerical results for the average longitudinal profile of energy within the sample, denoted $W_{\tau}(x)$, normalized so that $W_{\tau}(x=L)=\tau$ \cite{Choi12,Genack15}, in two ensembles consisting of $1000$ samples with channel number $N=80$, aspect ratio $2$, $\eta_L=5$, and the scaling conductance $g=N/\eta_L=16$. The mean free path is uniform in one ensemble of samples but varies with $x$ in the other. The profiles in two different ensembles overlap upon mapping $x$ to $\eta$$($$x$$)$.}

The invariance principle allows us to shift the focus from the surface to the interior of the sample and from homogeneous to inhomogeneous disorder. {We have demonstrated its application to intensity statistics in a general dimension and to controlling the spatial structure of transmission eigenchannels.} Our findings open a door to engineering the profiles of wave energy and its flow inside inhomogeneously disordered materials.

This work is supported by the National Science Foundation under grant number DMR/-BSF: 1609218 (A.Z.G.), by the National Natural Science Foundation of China under grant numbers 11535011, 11925507 and 11747601 (C.T.), and by MICINN (Spain) under the Project
number PGC2018-094684-B-C22 and Subprograma Estatal de Movilidad
2013-2016 under Project number PRX16/00166 (V.A.G).\\
\\
\noindent $^\dagger$These authors contribute equally to the work.\\
\noindent $^\ast$E-mail: azriel.genack@qc.cuny.edu\\
\noindent $^\ast$E-mail: gopar@unizar.es

\clearpage

\renewcommand{\thesection}{S\arabic{section}}
\renewcommand{\thesubsection}{\thesection.\arabic{subsection}}
\renewcommand{\theequation}{S\arabic{equation}}
\renewcommand{\thefigure}{S\arabic{figure}}

\setcounter{page}{1}
\setcounter{equation}{0}
\setcounter{figure}{0}

\begin{center}
{\bf Supplemental materials}\\
\end{center}

{This supplemental material is a self-contained
treatment of the (technical) experimental and theoretical discussion in the main paper and its relationship to the experiment.
For the convenience of the reader, all notation
and abbreviations are re-introduced and analytic
results given in the main paper are re-presented with links
to the main paper embedded into the
derivations. Thus readers who are interested only in the
technical part of this work may use this supplemental
material without going back and forth between
the main paper and this supplementary material.}

{The supplemental material is organized as follows. In Sec.~I, we present a full description of the experiments. In Sec.~II, we describe the supersymmetry (SUSY) theory in full and use this theory to derive the analytic results given in the paper. In Sec.~III, we give a full description of the random matrix theory (RMT) and use this theory to derive the analytic results given in the paper.}
{
\begin{center}
{\bf I. Description of experiments}\\
{\bf {\it 1. Experimental setup and methods of measurements}}
\end{center}
}
{Measurements are carried out with use of a
vector network analyzer in a single-mode rectangular copper waveguide containing
alternating ceramic layer and Teflon U-channel air spacer
layer. Each layer consists of one or more elements of the same material.
We use one type of ceramic element with thickness $6.6$ mm and
three types of spacers with thickness $1.27$, $2.55$ and $3.82$ cm.
Each elements are selected with probability of $1/2$ for the ceramic
element and $1/6$ for each thickness of the spacer.
%If we consider one pair of ceramic layer and spacer layer,
%the probability of having ceramic layer being $6.6n$mm ($n=1,2,3,\cdots$) is $\frac{1}{2^n}$,
%with an expectation value of $13.2$mm, and the probability of having spacer layer
%being $1.27n$cm ($n=1,2,3,\cdots$) is
%\begin{equation}\label{eq:S137}
%    \sum_{n_1,n_2,n_3}\left(\frac{1}{6}\right)^{n_1+n_2+n_3}\left(
%                                                              \begin{array}{c}
%                                                                n_1+n_2+n_3 \\
%                                                                n_1 \\
%                                                              \end{array}
%                                                            \right)
%\left(
%                                                              \begin{array}{c}
%                                                                n_2+n_3 \\
%                                                                n_2 \\
%                                                              \end{array}
%                                                            \right)
%\end{equation}
%provided $n_1+2n_2+3n_3=n$ $(n_1,n_2,n_3=0,1,2,\cdots)$,
%with an expectation value of $5.10$cm.
The copper waveguide with the cutoff frequency of
$6.56$ GHz contains the sample length $L=86.0$ cm. As a result,
the number of elements varies between configurations; the expected number of layers is around $30$.
This disorder is homogeneous and so gives rise to a linear falloff of the average of the logarithmic of intensity,
$\langle\ln I(x;x_0)\rangle$, within the sample. Spectra are taken for an ensemble of $100$
random configurations without a reflector and with a reflector at fixed relative
depths of $x_0/L=0,1/4,3/4,1$. The reflector is a thin copper plate covering 76$\%$
of the waveguide cross section with transmission coefficient $\Gamma=0.36$ in the empty waveguide.}

\begin{figure}[b]
\includegraphics[width=\columnwidth] {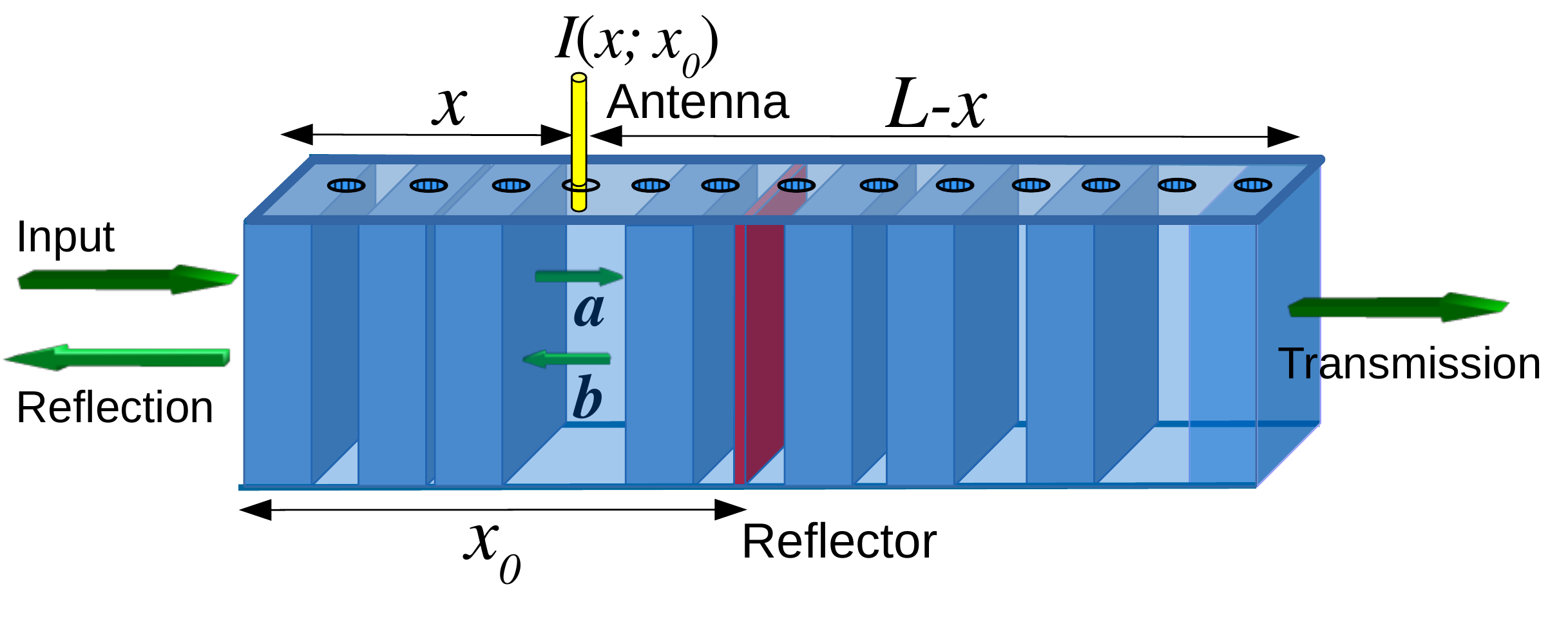}
\caption{Microwave radiation is launched into a random waveguide with a reflector at depth $x_0$. The detection antenna is inserted into holes in the waveguide to measure the field.}
\label{fig:1}
\end{figure}

{The wave is launched from the left into the random waveguide, where a reflector is placed at depth $x_0$, into the sample and detected by an antenna inserted sequentially into a series of holes of diameter $3.17$ mm spaced by $1$ cm along the waveguide (cf.~Fig.~\ref{fig:1}). Field spectra are taken over the frequency interval: $10.00-10.70$ GHz in which the waveguide supports a single mode. The antenna tip penetrates a depth $0.11$ mm below the bottom of the hole in the top plate of the waveguide. The incident intensity $I_0$ is found by fitting the expression for the intensity of counter-propagating waves: $I_0 [1+r^2+2r\cos(2kx+\varphi)]$ in a $4$-cm-long segment before the random sample to determine the incident field. Here $r$ is the magnitude of the reflected field, $k$ is the wavenumber determined from waveguide dispersion, and $\varphi$ is the phase. Measurements of intensity at a specific detector position, $I(x;x_0)$, are normalized by the measurements in an empty waveguide with absorbers at the ends of the waveguide so that the sensitivity is uniform for measurements at all positions.}

\begin{center}
{\bf {\it 2. Eliminating the impact of dissipation from microwave measurements}}
\end{center}

\begin{figure}[t]
\includegraphics[width=\columnwidth] {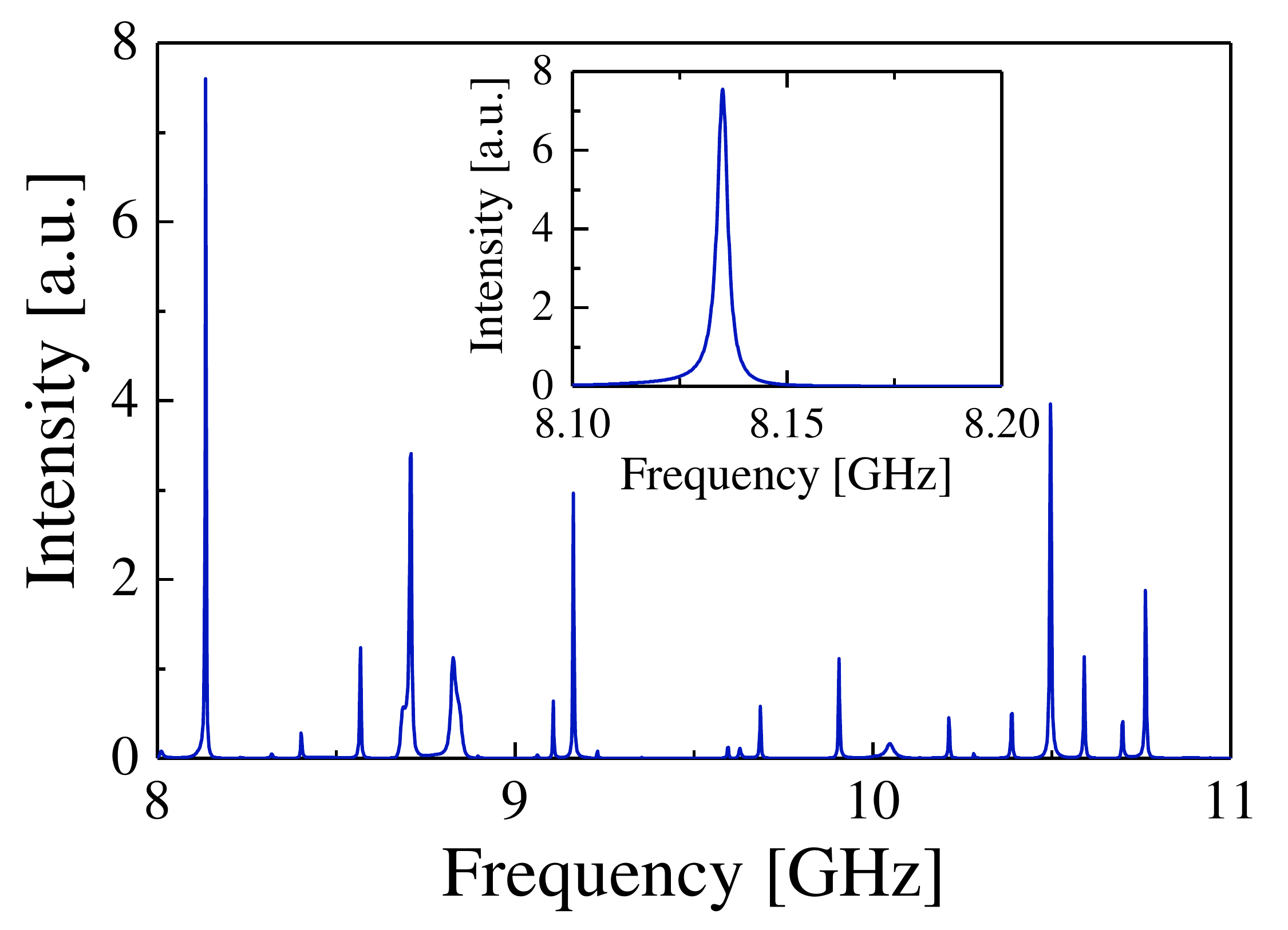}
\caption{Intensity spectrum in a random sample with a full reflector placed at the output and a $90\%$ reflector at the input. The spectrum of the narrowest peak in the wider spectrum is shown in the inset.}
\label{fig:S1}
\end{figure}

In this work, the statistics of wave propagation are discovered in microwave measurements. These measurements are used to confirm theoretical predictions of the equivalence of propagation statistics inside disordered samples in which a reflector is moved. To eliminate the impact of dissipation, which is primarily due to absorption, we first determine the decay rate of energy $\Gamma_a$ due to all sources
except for the leakage through the input and output boundaries of the sample. Sources of dissipation include absorption in the dielectric medium and in reflection from the copper waveguide, and leakage through the holes along the length of the sample.
We measure the average decay rate of the energy density at several points near the center of the sample in which leakage through the input and output boundaries of a sample is substantially blocked. A fully reflecting aluminum block is placed at the output of the sample and a copper reflector with approximately $10\%$ transmission is placed between the source and the sample input. In this way, coupling to the outside through the partial reflector is minimized, but the sample may still be excited by the antenna at the input side of the waveguide. An example of the intensity spectrum determined from measurements of the field in the center of the sample is shown in Fig.~\ref{fig:S1}. The peaks are narrowed by the reflectors at the ends of the sample, as can be seen by comparison with a typical spectrum with open boundaries. The narrowest lines  in the spectra correspond to modes that are strongly localized away from the partial reflector at the front of the sample. Thus the half-widths in angular frequency units of the narrowest lines in the spectrum are close to the rate of energy dissipation in the sample. In the spectrum shown in the inset of Fig.~\ref{fig:S1}, the narrowest peak at a frequency of $8.135\, {\rm GHz}$ corresponds to a mode strongly localized near the sample center which is not significantly broadened by leakage through the boundaries. This yields a decay rate of $9.3\times 10^{-3}\,{\rm ns}^{-1}$.

\begin{figure}[b]
\includegraphics[width=\columnwidth] {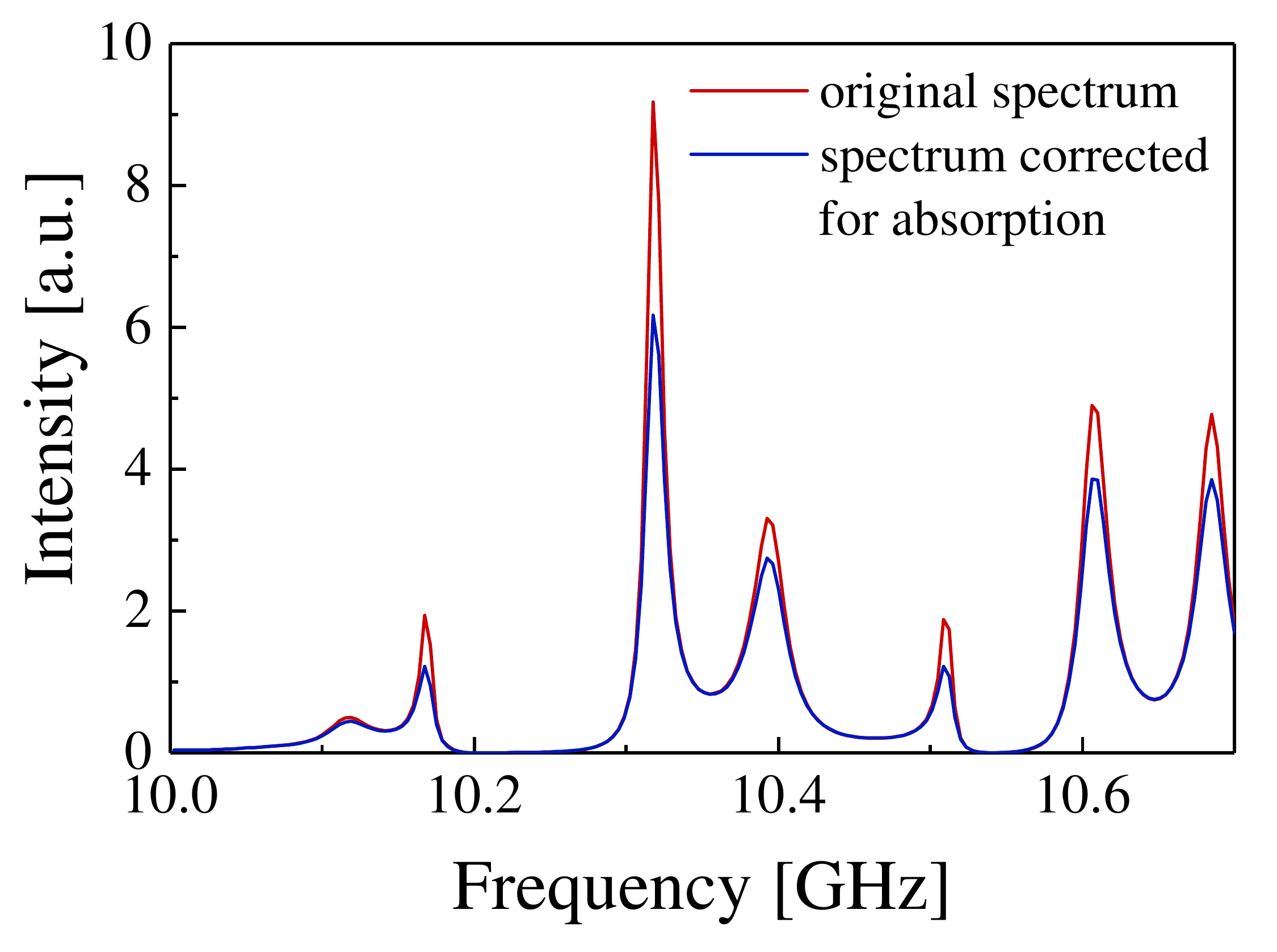}
\caption{Original spectrum at a point near the sample center taken from measurements and the spectrum after adjusting for the impact of absorption. }
\label{fig:S2}
\end{figure}

In order to find the spectrum in the absence of dissipation, we multiply the measured field spectrum in an open sample by the spectrum of a Gaussian function corresponding to the Fourier transform of the spectrum due to incident pulse. This spectrum is Fourier transformed into the time domain to yield the time evolution of the field inside the sample excited by the corresponding Gaussian pulse. This time varying field is then multiplied by $e^{\Gamma_a t/2}$ to give the time evolution of the field in a sample without dissipation. The modified time domain signal is then Fourier transformed back into the frequency domain to give the field spectra in which the influence of absorption is eliminated. These results are used to find the intensity statistics shown in the figures of the main text. This method has been shown in previous work \cite{Genack00SM} to give the probability distribution of intensity for a sample without absorption which is in agreement with predictions. The modification of the intensity spectrum when this method is applied to a point at the sample center of a typical disordered configuration is shown in Fig.~\ref{fig:S2}. This small change in the spectrum shows that the dissipation is weak in the present experiments.

The holes on the top of the copper waveguide of thickness $3.24$ mm have a diameter $3.17$ mm, which is $9.5$ times larger than the evanescent decay length of the microwave radiation within the hole, which is measured to be $0.34$ mm. Since the energy leakage through each hole is small and the holes cover a small fraction of the sides of the waveguide, this source of dissipation is negligible. We conclude that the decay of energy within the sample is largely due to absorption.

\begin{center}
{\bf II. Supersymmetry field theory for inhomogeneously disordered open media}
\end{center}

{In this section, we present a supersymmetry field theory for general inhomogeneously disordered open media. We use this theory to study the disorder average of a generic wave propagation quantity.}

{In most of this section, we consider quasi-$1$D and $1$D samples. Note that the local dimensionality of the quasi-1D sample is high. We first develop a SUSY theory for open media with an embedded reflector and use this to derive the general law of Eq.~(\ref{eq:1}) for a generic disorder-averaged quantity $O(x;x_0)$. For the convenience of the readers, we rewrite Eq.~(\ref{eq:1}) here:
\begin{eqnarray}\label{eq:S134}
    O(x;x_0)=\theta(x_0-x)O(x;L)+\theta(x-x_0)O(x;0),\quad
\end{eqnarray}
where $\theta$ is the Heaviside function. We further justify the generalized diffusion equation for the average intensity $\langle I(x;x_0)\rangle$ ($\langle\cdot\rangle$ the disorder average) and establish the scaling law of the position-dependent diffusion coefficient.} Then we extend SUSY to more general inhomogeneously disordered samples in which the strength of distributed scattering varies in space and multiple reflectors are embedded. This allows us to prove the generalized invariance principle and derive the analytic expressions for $\langle\ln I(x;x_0)\rangle$ and $\langle I(x;x_0)\rangle$ in these samples. {In the end of this section, we go beyond quasi-$1$D geometry and establish the invariance principle for high-dimensional disordered samples.}

\begin{center}
{\bf A. Field-theoretical expression for $\boldsymbol{O(x;x_0)}$}
\end{center}

We {first} develop the SUSY formalism for a quasi-$1$D random sample of length $L$ which, {for the sake of simplicity, is assumed to be} locally $2$D, and finally extrapolate the quasi-$1$D results to $1$D. The dielectric constant $\epsilon(\boldsymbol{r})$ exhibits independent Gaussian fluctuations around unity at every point $\boldsymbol{r}$$=$$($$x$$,$$y$$)$. A uniform reflective dielectric layer is placed at $x_0$. Wave propagation inside the medium is described by the retarded (advanced) Green function $G^{R(A)}$, defined from
\begin{equation}\label{eq:S93}
    \left(\nabla^2+\omega^2\left(\epsilon(\boldsymbol{r})\pm i0^+\right)\right)G^{R,A}(\boldsymbol{r},\boldsymbol{r}')=\delta(\boldsymbol{r}-\boldsymbol{r}').
\end{equation}
Recall that $\omega$ is the circular frequency.

Given a source profile $J(x)$ which is uniform in the $y$ direction, the average intensity integrated over the cross section is
\begin{equation}\label{eq:S71}
    \langle I(x;x_0)\rangle=\int\frac{d\omega}{2\pi}\int dx'{\cal Y}(x,x')J(x'),
\end{equation}
where
\begin{equation}\label{eq:58}
    {\cal Y}(x,x')\equiv \int\!\!\!\!\int dydy' \langle G^A(\boldsymbol{r},\boldsymbol{r}')G^R(\boldsymbol{r}',\boldsymbol{r})\rangle
\end{equation}
gives the spatial correlation between energy densities integrated over cross sections at $x$ and $x'$.

{Similar to the path-integral formalism in quantum mechanics, we can express a generic wave propagation quantity in terms of a path integral over a vector field. In the present context, because $G^{R}$ and $G^{A}$ have distinct analytic structures, this vector field carries a $2$-component index to accommodate these structures. In addition, it has another $2$-component index to accommodate the system's time-reversal symmetry. Most importantly, in order to make the normalization factor (the partition function) to be unity so that the subsequent disorder averaging can be readily carried out, we introduce fermion-boson doubling, which is the origin of SUSY. As such, the vector field is an $8$-component supervector field, denoted $\psi$. Now, since a local quantity such as $I$$($$x$$;$$x_0$$)$ can be expressed in terms of $G^{R(A)}$, upon performing the disorder average, one can express $O$$($$x$$;$$x_0$$)$ in terms of a functional integral over a supermatrix field $Q$ \cite{Efetov97}. Since the supermatrix $Q$ has the same structure as $\psi\otimes\psi^\dagger$, it is defined in the fermion-boson (`fb'), advanced-retarded (`ar') and time-reversal (`tr') sectors, each of which is a $2$$\times$$2$ block. The first sector accommodates the above doubling with complex and Grassmann variables as diagonal and non-diagonal entries, respectively; the second sector accommodates the distinct analytic structures of $G^{R(A)}$; the last sector accounts for time-reversal symmetry. The detailed structure of $Q$ depends on the system's symmetry. In general, we can parametrize $Q$ as
\begin{eqnarray}
\label{eq:S40}
Q&=&g\Lambda g^{-1},\nonumber\\
g&=&e^w,\quad \{w,\Lambda\}=0.
\end{eqnarray}
Here $g$ is an element of group ${\rm G}$ of supermatrices respecting the system's symmetry, and $g{\rm K}$ generates the coset space ${\rm G}/{\rm K}$, with the subgroup ${\rm K}$ consisting of all supermatrices commuting with the constant supermatrix $\Lambda$,
\begin{equation}\label{eq:S96}
    \Lambda=\mathbb{I}^{\rm fb}\otimes \sigma_3^{\rm ar}\otimes \mathbb{I}^{\rm tr}.
\end{equation}
In Eq.~(\ref{eq:S96}) $\sigma_i^{\rm X}$ ($i=1,2,3$ and $X$$=$`fb', `ar', `tr') and $\mathbb{I}^X$ are, respectively, the Pauli matrices and unit matrix in the X-sector.}

{Below let us consider a concrete example to show how this program is carried out in practice. Following Ref.~\cite{Tian08}, we can express ${\cal Y}$ as a functional integral over $Q(x)$,}
\begin{eqnarray}\label{eq:S7}
    {\cal Y}(x,x')&=&\int DQ_-DQ_+ e^{\gamma {\rm str}(Q_-Q_+)}\nonumber\\
    &\times&\int_{Q(0)=\Lambda}^{Q(x_0^-)=Q_-} D[Q]e^{-\frac{\xi}{8}\int_0^{x_0} dx{\rm str}(\partial_x Q)^2}\nonumber\\
    &\times&\int_{Q(x_0^+)=Q_+}^{Q(L)=\Lambda} D[Q]e^{-\frac{\xi}{8}\int_{x_0}^L dx{\rm str}(\partial_x Q)^2}\nonumber\\
    &\times&\frac{\pi\nu}{2^7}{\rm str}\left(A_+Q(x)A_-Q(x')\right),
\end{eqnarray}
with `str' being the supertrace,
\begin{equation}\label{eq:S95}
    A_\pm=(1-\sigma_3^{\rm fb})\otimes (1\pm\sigma_3^{\rm ar})\otimes (1-\sigma_3^{\rm tr}),
\end{equation}
and $\nu$ being the density of states per unit length. Here the action,
\begin{equation}\label{eq:S53}
\frac{\xi}{8}\int dx{\rm str}(\partial_x Q)^2,
\end{equation}
with $\xi$$=$$\pi$$\nu$$D_0$, reflects localization effects. Recall that $D_0$ is the Boltzmann diffusion coefficient. In Eq.~(\ref{eq:S7}), the tunneling action, $-$$\gamma$${\rm str}$$($$Q_-$$Q_+$$)$, appears, which couples the segment in front of and behind the reflector, where $Q_\pm$$\equiv$$Q(x_0^\pm)$ are the supermatrices on the two surfaces of the reflector.  Here the coupling constant $\gamma$ depends on the transmission coefficient $\Gamma$ of the reflector, the frequency $\omega$, and the cross sectional area $A$:
\begin{equation}\label{eq:S97}
    \gamma=\frac{\omega A}{4\pi} \frac{\Gamma}{2+\Gamma}.
\end{equation}
The first line of Eq.~(\ref{eq:S7}) corresponds to an integral over the supermatrices $Q_\pm$, with a normal measure $DQ_+DQ_-$. The second line corresponds to a functional integral over smooth field configurations in front of the reflector; which are subject to the boundary conditions: $Q$$($$0$$)$$=$$\Lambda$ and $Q$$($$x_0^-$$)$$=$$Q_-$ with a functional measure $D[Q(x)]$. The third line corresponds to a functional integral over smooth field configurations behind the reflector which are subject to the boundary conditions: $Q$$($$x_0^+$$)$$=$$Q_+$ and $Q$$($$L$$)$$=$$\Lambda$.

{To find the general field-theoretical expression of disorder-averaged quantities, we consider another example,} the transmission eigenvalue distribution $\rho(\tau)$$=$$\langle\sum_n\delta(\tau-\tau_n)\rangle$, for which the point of observation is $x$$=$$L$. Here $\tau_n$ is the $n$th transmission eigenvalue, namely the square of the singular value of the transmission matrix. By introducing the following quantity,
\begin{equation}\label{eq:S100}
    Z=\left\langle \frac{{\rm det}(1-\gamma_1\gamma_2
    \hat j \delta_{l}G^{A}\hat j \delta_{r}G^{R})}
    {{\rm det}(1-\zeta_1\zeta_2
    \hat j \delta_{l}G^{A}\hat j \delta_{r}G^{R})}\right\rangle,
\end{equation}
where $\hat j$ is the energy current operator and $\delta_{l(r)}$ restricts the spatial integral to the input (output), the distribution is expressed as \cite{Rejaei96SM}
\begin{equation}\label{eq:S94}
    \rho(\tau)=\frac{1}{\pi\tau}{\rm Im}\gamma_1\frac{\partial Z}{\partial \gamma_1}\bigg|_{\gamma_1\gamma_2=\zeta_1\zeta_2=\frac{1}{\tau+i0^+}}.
\end{equation}
Following the derivations of Ref.~\cite{Tian13bSM}, we find
\begin{eqnarray}\label{eq:S101}
    Z&=&\int DQ_-DQ_+ e^{\gamma {\rm str}(Q_-Q_+)}\nonumber\\
    &\times&\int_{Q(0)=\Lambda}^{Q(x_0^-)=Q_-} D[Q]e^{-\frac{\xi}{8}\int_0^{x_0} dx{\rm str}(\partial_x Q)^2}\nonumber\\
    &\times&\int_{Q(x_0^+)=Q_+}^{Q(L)=Q_L} D[Q]e^{-\frac{\xi}{8}\int_{x_0}^L dx{\rm str}(\partial_x Q)^2},
\end{eqnarray}
where $Q_L$ is determined completely by $\gamma_{1,2}$ and $\zeta_{1,2}$.

Comparing Eqs.~(\ref{eq:S7}) and (\ref{eq:S101}), we find that the ensemble average of a generic quantity measured at $x$, which is represented by $O(x;x_0)$, can be expressed as
\begin{eqnarray}\label{eq:S4}
    &&O(x;x_0)\equiv\int DQ_-DQ_+ e^{\gamma {\rm str}(Q_-Q_+)}\nonumber\\
    &\times&\int_{Q(0)=\Lambda}^{Q(x_0^-)=Q_-} D[Q]e^{-\frac{\xi}{8}\int_0^{x_0} dx{\rm str}(\partial_x Q)^2}\nonumber\\
    &\times&\int_{Q(x_0^+)=Q_+}^{Q(L)=Q_L} D[Q]e^{-\frac{\xi}{8}\int_{x_0}^{L} dx{\rm str}(\partial_x Q)^2}M_O[Q(x)],\quad
\end{eqnarray}
and only $M_O[Q(x)]$ and the boundary condition, $Q$$($$L$$)$$=$$Q_L$, are specific to $O$. Similar to Eq.~(\ref{eq:S7}), the first line of Eq.~(\ref{eq:S4}) corresponds to an integral over the supermatrices $Q_\pm$ at the two surfaces of the reflector, the second to a functional integral over smooth field configurations in front of the reflector, which are subject to the boundary conditions: $Q$$($$0$$)$$=$$\Lambda$ and $Q$$($$x_0^-$$)$$=$$Q_-$, and the third to a functional integral over smooth field configurations behind the reflector, which is subject to the boundary conditions: $Q$$($$x_0^+$$)$$=$$Q_+$ and $Q$$($$L$$)$$=$$Q_L$.

{
\begin{center}
{\bf B. Proof of invariance principle for inhomogeneity arising from an embedded reflector}
\end{center}
}
We are now ready to prove Eq.~(\ref{eq:S134}). The proof includes three steps. In the first, we introduce the heat kernel. In the second, we prove an identity for the heat kernel. With these preparations, we complete the proof of Eq.~(\ref{eq:S134}) in the third step.

\begin{center}
{\bf {\it 1. Step I}}
\end{center}

To demonstrate the basic idea, we consider, for the moment, samples without a reflector and thus the parameter $x_0$ in $O(x;x_0)$ is suppressed. In this case, Eq.~(\ref{eq:S4}) is simplified to
\begin{eqnarray}\label{eq:S1}
    O(x)=\int_{Q(0)=\Lambda}^{Q(L)=Q_L} D[Q]e^{-\frac{\xi}{8}\int_0^{L} dx{\rm str}(\partial_x Q)^2}M_O[Q(x)],\quad\quad
\end{eqnarray}
were the functional integral is over smooth field configurations subject to the boundary conditions: $Q$$($$0$$)$$=$$\Lambda$ and $Q$$($$L$$)$$=$$Q$. For convenience below, we rescale $x$ by $4\xi$. The common structure of Eq.~(\ref{eq:S1}) motivates us to introduce the heat kernel \cite{Zirnbauer94SM,Helgason78SM}, defined as
\begin{equation}\label{eq:S3}
    W(Q,Q';\tilde L)\equiv \int_{Q(0)=Q'}^{Q(\tilde L)=Q} D[Q]e^{-\frac{1}{8}\int_0^{\tilde L} dx{\rm str}(\partial_x Q)^2},
\end{equation}
where $\tilde L$$=$$L$$/$$($$4$$\xi$$)$. The heat kernel satisfies
\begin{eqnarray}
\label{eq:S31}
  \left(\partial_{\tilde L}-\Delta_Q\right)W(Q,Q';\tilde L)=\delta(Q,Q')\delta(\tilde L),
\end{eqnarray}
where $\Delta_Q$ is the Laplacian in  $Q$ space. Comparing Eq.~(\ref{eq:S31}) with the normal diffusion equation describing the macroscopic motion of Brownian particles, we find that $\tilde L$ and $Q$ play the roles of ``time'' and ``coordinate'', respectively, and $W$ mimics a ``probability distribution''. As such, the action ${\rm str}$$($$\partial_x$$Q$$)^2$ mimics the kinetic energy of a Brownian particle. Using Eq.~(\ref{eq:S3}), we can rewrite Eq.~(\ref{eq:S1}) as
\begin{eqnarray}
    O(x)&=&\int\!\! DQ(x)W(Q_L,Q(x);\tilde L-x))W(Q(x),\Lambda;x)\nonumber\\
&\times&M_O[Q(x)].
\label{eq:S2}
\end{eqnarray}
This expresses $O(x)$ in terms of the heat kernel and reduces the functional integral over supermatrix field configurations to the integral over a single matrix $Q(x)$.

\begin{figure}[t]
\includegraphics[width=\columnwidth] {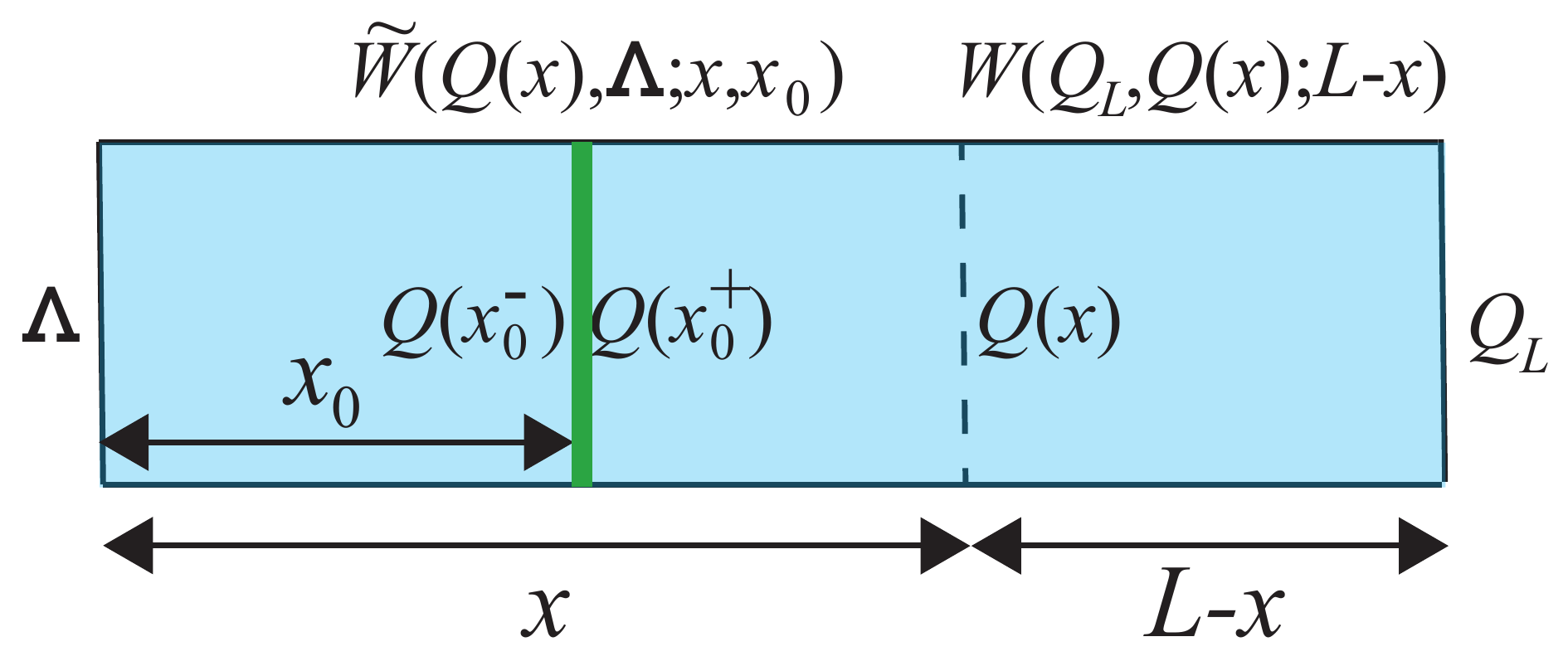}
\caption{Schematic representation of Eq.~(\ref{eq:S48}) for $x$$>$$x_0$.}
\label{fig:6}
\end{figure}

We then generalize this idea to samples with a reflector at depth $x_0$ rescaled by $4\xi$ and express $O(x;x_0)$ in terms of the heat kernel. We introduce the following quantity,
\begin{eqnarray}\label{eq:S34}
    &&{\tilde W}(Q,Q';x_1,x_2)\equiv\int DQ_-DQ_+ e^{\gamma {\rm str}(Q_-Q_+)}\nonumber\\
    &\times&\int_{Q(0)=Q'}^{Q(x_2^-)=Q_-} D[Q]    e^{-\frac{1}{8}\int_0^{x_2} dx{\rm str}(\partial_x Q)^2}\nonumber\\
    &\times&\int_{Q(x_2^+)=Q_+}^{Q(x_1)=Q} D[Q]e^{-\frac{1}{8}\int_{x_2}^{x_1} dx{\rm str}(\partial_x Q)^2}
\end{eqnarray}
for $x_1$$>$$x_2$. With the substitution of Eq.~(\ref{eq:S3}), we obtain
\begin{eqnarray}\label{eq:S35}
    &&{\tilde W}(Q,Q';x_1,x_2)\equiv\int DQ_-DQ_+ e^{\gamma {\rm str}(Q_-Q_+)}\nonumber\\
    &\times& W(Q,Q_+;x_1-x_2)W(Q_-,Q';x_2).
\end{eqnarray}
Combining Eqs.~(\ref{eq:S4}), (\ref{eq:S34}) and (\ref{eq:S35}) gives
\begin{eqnarray}
    O(x;x_0)=\int DQ(x)M_O[Q(x)]\qquad\qquad\qquad\qquad\label{eq:S48}\\
    \times\Big\{\begin{array}{c}
                    W(Q_L,Q(x);\tilde L-x){\tilde W}(Q(x),\Lambda;x,x_0),\, x>x_0\\
                    {\tilde W}(Q_L,Q(x);\tilde L-x,x_0-x)W(Q(x),\Lambda;x),\, x<x_0\,\,
                  \end{array}
    \nonumber
\end{eqnarray}{namely Eq.~(\ref{eq:48}),} which expresses $O$ in terms of the heat kernel (Fig.~\ref{fig:6}). {Note that Eqs.~(\ref{eq:S4}) and (\ref{eq:S48}) are equivalent.}

\begin{center}
{\bf {\it 2. Step II}}
\end{center}

Taking the derivative with respect to $x_2$ on both sides of Eq.~(\ref{eq:S35}) gives
\begin{eqnarray}\label{eq:S37}
    \partial_{x_2}{\tilde W}(Q,Q';x_1,x_2)=\int DQ_-DQ_+ e^{\gamma {\rm str}(Q_-Q_+)}\nonumber\\
    \times\big(\left(\partial_{x_2}W(Q_-,Q';x_2)\right)W(Q,Q_+;x_1-x_2)\quad\quad\nonumber\\
    +W(Q_-,Q';x_2)\left(\partial_{x_2}W(Q,Q_+;x_1-x_2)\right)\big).\quad
\end{eqnarray}
With the help of Eq.~(\ref{eq:S31}), this reduces to
\begin{eqnarray}\label{eq:S38}
    \partial_{x_2}{\tilde W}(Q,Q';x_1,x_2)=\int DQ_-DQ_+ e^{\gamma {\rm str}(Q_-Q_+)}\nonumber\\
    \times\big(\left(\Delta_{Q_-}W(Q_-,Q';x_2)\right)W(Q,Q_+;x_1-x_2)\quad\quad\nonumber\\
    -W(Q_-,Q';x_2)\left(\Delta_{Q_+}W(Q,Q_+;x_1-x_2)\right)\big).\quad
\end{eqnarray}
Using integration by parts in the superanalysis \cite{Brezin87SM}, Eq.~(\ref{eq:S38}) can be rewritten as
\begin{eqnarray}\label{eq:S39}
    &&\partial_{x_2}{\tilde W}(Q,Q';x_1,x_2)\nonumber\\
    &=&\int DQ_-DQ_+W(Q_-,Q';x_2)W(Q,Q_+;x_1-x_2)\nonumber\\
    &\times&\left(\Delta_{Q_-}-\Delta_{Q_+}\right) e^{\gamma {\rm str}(Q_-Q_+)}.
\end{eqnarray}
Below, we show that factor given in the last line of Eq.~(\ref{eq:S39}) vanishes.

For this purpose, we  parametrize $Q_\pm$ as
\begin{equation}\label{eq:S41}
    Q_\pm=g_\pm\Lambda g_\pm^{-1},\quad g_\pm=e^{w_\pm},\quad \{w_\pm,\Lambda\}=0
\end{equation}
using Eq.~(\ref{eq:S40}). With this substitution, we obtain
\begin{eqnarray}
\label{eq:S42}
\Delta_{Q_+}e^{\gamma {\rm str}(Q_-Q_+)}=\Delta_{g_+{\rm K}}e^{\gamma {\rm str}(g_-^{-1}g_+\Lambda g_+^{-1}g_-\Lambda)}.\quad
\end{eqnarray}
Because the left group translation,
\begin{equation}\label{eq:S56}
    L_{h}: {\cal F}(g)\rightarrow L_{h}{\cal F}(g)={\cal F}(hg),
\end{equation}
with ${\cal F}$ a generic function over the coset space, commutes with the Laplacian, letting ${\cal F}$$($$g$$)$$=$$e^{\gamma {\rm str}(g\Lambda g^{-1}\Lambda)}$ and $g$$=$$g_+$$,$$h$$=$$g_-^{-1}$, we have
\begin{eqnarray}
\Delta_{Q_+}e^{\gamma {\rm str}(Q_-Q_+)}&=&\Delta_{\tilde g{\rm K}}e^{\gamma {\rm str}(\tilde g\Lambda \tilde g^{-1}\Lambda)},\label{eq:S43}\\
\tilde g&\equiv&g_-^{-1}g_+.\nonumber
\end{eqnarray}
Note that $\tilde g{\rm K}\in {\rm G}/{\rm K}$ and thus can be parametrized in the same way as Eq.~(\ref{eq:S40}). That is, we have $\tilde g=e^{\tilde w}$ with $\{\tilde w,\Lambda\}=0$. The right-hand side of Eq.~(\ref{eq:S43}) can be calculated by using the definition of the Laplacian on the coset space \cite{Zirnbauer94SM,Helgason78SM}, giving
\begin{eqnarray}
\label{eq:S44}
&&\Delta_{\tilde g{\rm K}}e^{\gamma {\rm str}(\tilde g\Lambda \tilde g^{-1}\Lambda)}\nonumber\\
&=&\sum_{i,j}g^{ij}\partial_{x_i}\partial_{x_j}e^{\gamma {\rm str}(\tilde ge^x\Lambda (\tilde ge^x)^{-1}\Lambda)}\big|_{x=0},
\end{eqnarray}
where $\{g^{ij}\}$ is the metric tensor on curved coset space and $\{x,\Lambda\}=0$. With the help of the anticommutation relation between $x$ (or $\tilde w$) and $\Lambda$ and Eq.~(\ref{eq:S43}), we have
\begin{eqnarray}
\label{eq:S45}
\Delta_{Q_+}e^{\gamma {\rm str}(Q_-Q_+)}=\sum_{i,j}g^{ij}\partial_{x_i}\partial_{x_j}e^{\gamma {\rm str}(e^{2\tilde w}e^{2x})}\big|_{x=0}.\nonumber\\
\end{eqnarray}
Similarly, we have
\begin{eqnarray}
\label{eq:S46}
\Delta_{Q_-}e^{\gamma {\rm str}(Q_-Q_+)}&=&\Delta_{\tilde g^{-1}{\rm K}}e^{\gamma {\rm str}(\tilde g^{-1}\Lambda\tilde g\Lambda)}\nonumber\\
&=&\sum_{i,j}g^{ij}\partial_{x_i}\partial_{x_j}e^{\gamma {\rm str}(e^{2\tilde w}e^{-2x})}\big|_{x=0}.\nonumber\\
\end{eqnarray}
Comparing Eqs.~(\ref{eq:S45}) and (\ref{eq:S46}) and taking into account the quadratic property of the Laplacian, we find that
\begin{equation}\label{eq:S47}
    \Delta_{Q_+}e^{\gamma {\rm str}(Q_-Q_+)}=\Delta_{Q_-}e^{\gamma {\rm str}(Q_-Q_+)}.
\end{equation}
Combining this with Eq.~(\ref{eq:S39}) proves the following identity,
\begin{eqnarray}\label{eq:36}
    \partial_{x_2}{\tilde W}(Q,Q';x_1,x_2)=0.
\end{eqnarray}
Recall that here $x_1>x_2$.

\begin{center}
{\bf {\it 3. Step III}}
\end{center}

Consider $O$$($$x$$;$$x_0$$)$ with $x_0$$<$$x$. Equation (\ref{eq:S48}) gives
\begin{eqnarray}\label{eq:S103}
    O(x;x_0<x)=\int DQ(x)M_O[Q(x)]\quad\quad\quad\quad\quad\nonumber\\
    \times{\tilde W}(Q(x),\Lambda;x,x_0)W(Q_L,Q(x);\tilde L-x).
\end{eqnarray}
Taking the derivative with respect to $x_0$ and using the identity Eq.~(\ref{eq:36}), we have
\begin{eqnarray}\label{eq:S5}
    \partial_{x_0}O(x;x_0<x)=0,
\end{eqnarray}
which gives
\begin{eqnarray}\label{eq:S49}
    O(x;x_0<x)=O(x;0).
\end{eqnarray}
Similarly, for $x<x_0$, Eq.~(\ref{eq:S48}) gives
\begin{eqnarray}\label{eq:S50}
    O(x;x_0>x)=\int DQ(x)M_O[Q(x)]\quad\quad\quad\quad\quad\quad\nonumber\\
    \times W(Q(x),\Lambda;x){\tilde W}(Q_L,Q(x);\tilde L-x,x_0-x).\quad\quad
\end{eqnarray}
Taking the derivative with respect to $x_0$ and using the identity Eq.~(\ref{eq:36}), we have
\begin{eqnarray}\label{eq:S51}
    \partial_{x_0}O(x;x_0>x)=0,
\end{eqnarray}
which gives
\begin{eqnarray}\label{eq:S52}
    O(x;x_0>x)=O(x;L).
\end{eqnarray}
From Eqs.~(\ref{eq:S49}) and (\ref{eq:S52}) we obtain Eq.~(\ref{eq:S134}).

\begin{center}
{\bf C. Generalized diffusion equation}
\end{center}

We now calculate ${\cal Y}$$($$x$$,$$x'$$)$ by using its field-theoretical expression given by Eq.~(\ref{eq:S7}). We generalize the scheme which was developed for samples without a reflector \cite{Tian08,Tian10,Tian13aSM}. Below the rescaling of $x,x_0,L$ by $4\xi$ is undone. First of all, the integral over supermatrices $Q_\pm$ is dominated by those matrices of $Q_\pm$ so that the total action is invariant under an infinitesimal rotation at $x_0^\pm$,
\begin{eqnarray}
Q_\pm \rightarrow e^{-\delta R_\pm}Q_\pm e^{\delta R_\pm}.
\label{eq:S13}
\end{eqnarray}
Substituting this into the action and demanding that the term linear in $\delta R_\pm$ vanishes, we obtain
\begin{eqnarray}
  4\zeta Q_+\partial_x Q_+ - [Q_+,Q_-] &=& 0, \label{eq:S14}\\
  4\zeta Q_-\partial_x Q_- - [Q_+,Q_-] &=& 0, \label{eq:S15}
\end{eqnarray}
where $\zeta
%(\Gamma)
$$=$$\xi/(4\gamma)
%\frac{\sqrt{\pi} \ell}{d}\frac{\Gamma((d+1)/2)}{\Gamma(d/2)}\frac{2-\Gamma}{\Gamma}
$. Subtracting these two equations gives
\begin{eqnarray}
  Q_+\partial_x Q_+=Q_-\partial_x Q_-,
\label{eq:S16}
\end{eqnarray}
which implies that the energy flux through the reflector is conserved. Thus among the three constraints of Eqs.~(\ref{eq:S14}), (\ref{eq:S15}) and (\ref{eq:S16}), only two are independent. Below we choose Eqs.~(\ref{eq:S14}) and (\ref{eq:S16}).

Next, we consider weakly localized samples ($L$$\ll$$\xi$). For these samples, Eq.~(\ref{eq:S7}) can be calculated perturbatively. To facilitate this scheme, we perturb $Q(x)$ near $\Lambda$, by expanding $Q(x)$ in terms of $w(x)$ as
\begin{eqnarray}
\label{eq:S6}
Q(x)=\sum_{n=0}^\infty \frac{2^n}{n!}w^n(x)\Lambda
\end{eqnarray}
with use of Eq.~(\ref{eq:S40}). Because $Q(x)$ is fixed to be $\Lambda$ at $x=0,L$, we have the following boundary constraint,
\begin{eqnarray}
\label{eq:S8}
w(0)=w(L)=0.
\end{eqnarray}
Moreover, by substituting this expansion into Eqs.~(\ref{eq:S14}) and (\ref{eq:S16}), we obtain the following constraints for $w(x)$ on both sides of the reflector,
\begin{eqnarray}
  \zeta \partial_x w(x_0^+) - (w(x_0^+)-w(x_0^-))=0,\label{eq:S10}\\
  \partial_x w(x_0^+)=\partial_x w(x_0^-).\quad\quad\quad\quad\quad\label{eq:S11}
\label{eq:S12}
\end{eqnarray}
By keeping the leading order expansion of Eq.~(\ref{eq:S7}) in $w$, we find that the result, denoted ${\cal Y}_0(x,x')$, is the fundamental solution of the normal diffusion equation,
\begin{eqnarray}
  -D_0\partial_x^2 {\cal Y}_0(x,x')=\delta(x-x'),
\label{eq:S19}
\end{eqnarray}
and satisfies the following constraints:
\begin{eqnarray}
  \zeta\partial_x {\cal Y}_0(x_0^+,x')={\cal Y}_0(x_0^+,x')-{\cal Y}_0(x_0^-,x'),
\label{eq:S20}
\end{eqnarray}
\begin{eqnarray}
\partial_x {\cal Y}_0(x_0^+,x')=\partial_x {\cal Y}_0(x_0^-,x'),
\label{eq:S21}
\end{eqnarray}
\begin{eqnarray}
{\cal Y}_0(x=0,x')={\cal Y}_0(x=L,x')=0.
\label{eq:S22}
\end{eqnarray}
Solving this equation, we find that for $x'<x_0$
\begin{eqnarray}\label{eq:S23}
    {\cal Y}_0(x,x')=\left\{\begin{array}{cc}
                              \frac{x(L+\zeta-x')}{D_0(L+\zeta)} & {\rm for}\, 0\leq x\leq x' \\
                              \frac{x'(L+\zeta-x)}{D_0(L+\zeta)} & {\rm for}\, x'\leq x<x_0 \\
                              \frac{x'(L-x)}{D_0(L+\zeta)} & {\rm for}\, x_0<x\leq L
                            \end{array}
    \right.,
\end{eqnarray}
while for $x'>x_0$
\begin{eqnarray}\label{eq:S24}
    {\cal Y}_0(x,x')=\left\{\begin{array}{cc}
                              \frac{x(L-x')}{D_0(L+\zeta)} & {\rm for}\, 0\leq x<x_0 \\
                              \frac{(x+\zeta)(L-x')}{D_0(L+\zeta)} & {\rm for}\, x_0<x\leq x' \\
                              \frac{(x'+\zeta)(L-x)}{D_0(L+\zeta)} & {\rm for}\, x'\leq x\leq L
                            \end{array}
    \right..
\end{eqnarray}
Thus, up to second order in $w$, Eq.~(\ref{eq:S7}) gives the same result as the solution to the normal diffusion equation, with diffusion coefficient $D_0$.

We proceed to keep higher order terms in the $w$ expansion of Eq.~(\ref{eq:S7}). We find that ${\cal Y}$$=$${\cal Y}_0$$+$$\delta {\cal Y}$, where
\begin{eqnarray}
\label{eq:S25}
  \delta {\cal Y}(x,x')&=&-\int_0^L dx_1 \delta D(x_1;x_0)\nonumber\\
  &\times&\left(\partial_{x_1}{\cal Y}_0(x,x_1)\right)\left(\partial_{x_1}{\cal Y}_0(x_1,x')\right),
\end{eqnarray}
where
\begin{eqnarray}
\frac{\delta D(x;x_0)}{D_0}&=&-C_1\frac{{\cal Y}_0(x,x)}{\pi\nu}-\frac{C_2}{2}\left(\frac{{\cal Y}_0(x,x)}{\pi\nu}\right)^2\nonumber\\
&+&{\cal O}\left(\left(\frac{{\cal Y}_0(x,x)}{\pi\nu}\right)^3\right),
\label{eq:S26}
\end{eqnarray}
with the coefficients $C_1=1$ and $C_2=0$. The first and second terms correspond, respectively, to the leading and subleading weak localization corrections to $D_0$. With the help of Eqs.~(\ref{eq:S23}) and (\ref{eq:S24}), this reduces to
\begin{eqnarray}
\frac{\delta D(x;x_0)}{D_0}=-C_1\lambda(x;x_0)-\frac{C_2}{2}\lambda(x;x_0)
+{\cal O}\left(\lambda(x;x_0)^3\right),\nonumber\\
\label{eq:S27}\\
\lambda(x;x_0)=\frac{{\cal Y}_0(x,x)}{\pi\nu}=\left\{\begin{array}{ll}
                      \frac{(x+\zeta)(L-x)}{\xi (L+\zeta)} & {\rm for}\,x>x_0  \\
                      \frac{x(L+\zeta-x)}{\xi (L+\zeta)} & {\rm for}\,x<x_0
                    \end{array}.\quad\quad\quad\right.
\label{eq:S28}
\end{eqnarray}
With the substitution of Eq.~(\ref{eq:S28}), Eq.~(\ref{eq:S25}) can be rewritten as
\begin{eqnarray}
  -\partial_x D(x;x_0)\partial_x {\cal Y}(x,x')=\delta(x-x'),
\label{eq:S29}
\end{eqnarray}
where
\begin{eqnarray}
&&\frac{D(x;x_0)}{D_0}=\frac{D_0+\delta D(x;x_0)}{D_0}\nonumber\\
&=&1-C_1\lambda(x;x_0)-\frac{C_2}{2}\lambda(x;x_0)
+{\cal O}\left(\lambda(x;x_0)^3\right).\quad
\label{eq:S30}
\end{eqnarray}
Comparing this with Eq.~(\ref{eq:S19}), we see that $D_0$ is renormalized to $D(x;x_0)$ due to localization effects, and that the renormalization corrections are position dependent. More importantly, the position-dependent diffusion coefficient $D(x;x_0)$, when rescaled by $D_0$, depends on the observation point $x$, the reflector location $x_0$, the reflector transmission coefficient $\Gamma$ and the mean free path $\ell$ only via the scaling factor $\lambda(x;x_0)$. In addition, expansion coefficients such as $C_1$, $C_2$ are the same for samples with and without a reflector. This is equivalent to the existence of a scaling function $D_\infty(\lambda)$, such that
\begin{eqnarray}
\frac{D(x;x_0)}{D_0}=D_\infty(\lambda(x;x_0))
\label{eq:S32}
\end{eqnarray}
and $D_\infty(\lambda)$ is the same for samples with and without a reflector, and so is universal with respect to the explicit form of $\lambda$. By using Eqs.~(\ref{eq:S28}) and (\ref{eq:S32}), we find that
\begin{eqnarray}\label{eq:S58}
    D(x;x_0)=\theta(x-x_0)D(x;0)+\theta(x_0-x)D(x;L).\quad
\end{eqnarray}
This shows that $D(x;x_0)$ obeys the general law of Eq.~(\ref{eq:S134}).

Equation (\ref{eq:S30}) further gives the explicit form of the scaling function,
\begin{eqnarray}
\label{eq:S33}
D_\infty(\lambda)=1-C_1\lambda-\frac{C_2}{2}\lambda^2+{\cal O}\left(\lambda^3\right),\,{\rm for}\, \lambda\ll 1.\quad
\end{eqnarray}
This justifies the generalized diffusion equation and the scaling law of $D(x;x_0)$. So far, we have considered systems with time reversal symmetry. When time reversal symmetry is broken, the results described by Eqs.~(\ref{eq:S29})-(\ref{eq:S33}) are unchanged, except that the coefficients in Eq.~(\ref{eq:S33}) are changed to $C_1=0$ and $C_2=1$. That $C_1$ vanishes reflects the well-known result, that, for systems with broken time-reversal symmetry, the one-loop weak localization correction disappears. Instead, in this case, the leading interference effects start from the two loop terms. Interestingly, these include both localization and delocalization corrections, with the former dominating over the latter \cite{Tian13aSM,Tian05SM,Garreau18SM}.

To find the explicit form of $D_\infty(\lambda)$ for $\lambda\gg 1$, we exploit its universality and consider a semi-infinite sample ($L=\infty$) without a reflector. Moreover, because in quasi-$1$D or $1$D localization physics, results in the presence or absence of time-reversal symmetry (namely, orthogonal or unitary symmetry) differ only in the numerical factor in the localization length \cite{Efetov97,Kravtsov05SM}, we consider the simplest case of a sample with broken time-reversal symmetry. In this case, we calculate Eq.~(\ref{eq:S7}) exactly and find (below the argument $x_0$ is suppressed when the reflector is absent.)
\begin{equation}\label{eq:S54}
    D(x)=D_0e^{-x/\xi+{\cal O}(\ln (x/\xi))},\,{\rm for}\, x/\xi\gg 1.
\end{equation}
On the other hand, using Eq.~(\ref{eq:S28}), we find that $\lambda$$($$x$$)$$=$$x$$/$$\xi$ for semi-infinite samples without a reflector. Combining this with Eq.~(\ref{eq:S54}), we obtain
\begin{eqnarray}
\label{eq:S9}
D_\infty(\lambda)=e^{-\lambda+{\cal O}(\ln\lambda)},\,{\rm for}\,\lambda\gg 1.
\end{eqnarray}
Note that this result is asymptotic and thus Eq.~(\ref{eq:S33}) is not its Taylor expansion. This can also be seen noting that $C_1$ in the perturbative expansion Eq.~(\ref{eq:S33}) vanishes for broken time-reversal symmetry.

In $1$D, there is no weakly localized regime, since the localization length and the mean free path are of the same order. Thus, $D_\infty(\lambda)\approx e^{-\lambda}$ for this case, and the localization length $\xi$ appearing in the scaling factor $\lambda(x;x_0)$ given in Eq.~(\ref{eq:S28}) is replaced by a length, $\xi_{\rm 1D}$, which is of the order of the mean free path $\ell$. The precise value of $\xi_{\rm 1D}$ can be found as follows. Consider a strongly localized sample without a reflector for which $L\gg \xi_{\rm 1D}$. Because the average conductance (which is the average transmission coefficient) $\langle G\rangle$ is given by \cite{Tiggelen00,Tian10}
\begin{equation}\label{eq:S55}
    \langle G\rangle=\left(\int_0^L\frac{dx}{D(x)}\right)^{-1},
\end{equation}
by substituting $D(x)=D_0e^{-\frac{x(L-x)}{\xi_{\rm 1D}L}}$, we obtain
\begin{equation}\label{eq:S57}
    \langle G\rangle\sim e^{-L/(4\xi_{\rm 1D})}.
\end{equation}
On the other hand, it is well known from RMT \cite{Beenakker97SM,Gertsenshtein59} that $\langle$$G$$\rangle$$\sim$$e^{-L/(4\ell)}$ for a strongly localized $1$D sample. Comparing this result with Eq.~(\ref{eq:S57}), we obtain
\begin{equation}\label{eq:S59}
    \xi_{\rm 1D}=\ell.
\end{equation}
Thus the scaling factor given by Eq.~(\ref{eq:S28}) reduces to
\begin{eqnarray}
\label{eq:S60}
\lambda(x;x_0)=\left\{\begin{array}{ll}
                      \frac{(x+\zeta)(L-x)}{\ell (L+\zeta)} & {\rm for}\,x>x_0  \\
                      \frac{x(L+\zeta-x)}{\ell (L+\zeta)} & {\rm for}\,x<x_0
                    \end{array}\,\,\,\right.
\end{eqnarray}
for $1$D samples with a reflector placed at $x_0$. {From this Eq.~(\ref{eq:2}), namely,
\begin{eqnarray}
\label{eq:S140}
D(x;x_0)=D_0e^{-\lambda(x;x_0)}
\end{eqnarray}follows.}

\begin{center}
{\bf D. Intensity profile $\boldsymbol{\langle I(x;x_0)\rangle}$ in $\boldsymbol{1}$D}
\end{center}

In experiments, the source is at the boundaries of the sample. Because of this, we can use Eqs.~(\ref{eq:S71}) and (\ref{eq:S29}) to obtain Eq.~(\ref{eq:3}), namely, \begin{eqnarray}
  -\partial_x D(x;x_0)\partial_x \langle I(x;x_0)\rangle=0
\label{eq:S104}
\end{eqnarray}
within the medium. When implementing this homogeneous generalized diffusion equation with appropriate conditions at $x=0,L,x_0$ we can find the spatial profile $\langle I(x;x_0)\rangle$. Thus the profile is controlled by the position-dependent diffusion coefficient $D(x;x_0)$.

\begin{center}
{\bf {\it 1. Symmetry of intensity profile}}
\end{center}

We first discuss the symmetry of the average intensity profile, which is independent of the explicit form of $D(x;x_0)$. From Eq.~(\ref{eq:S104}), we find that the Fick current, $-$$D$$(x;x_0)$$\partial_{x}$$\langle I(x;x_0)\rangle$, is independent of $x$ in the segment in front of and behind the reflector. Moreover, the Fick current in the two segments must be the same. As a result,
\begin{eqnarray}
  &&-D(x;x_0)\partial_{x} \langle I(x;x_0)\rangle|_{x<x_0}\nonumber\\
  &=&-D(x;x_0)\partial_{x} \langle I(x;x_0)\rangle|_{x>x_0}.
\label{eq:S72}
\end{eqnarray}
On the other hand, the average intensity obeys the general law of Eq.~(\ref{eq:S134}). Thus Eq.~(\ref{eq:28}), namely,
\begin{equation}\label{eq:S73}
\langle I(x;x_0)\rangle=\theta(x-x_0)\langle I(x;0)\rangle+\theta(x_0-x)\langle I(x;L)\rangle
\end{equation}
follows. Combining Eq.~(\ref{eq:S72}) with Eqs.~(\ref{eq:S58}) and (\ref{eq:S73}) gives
\begin{eqnarray}
  &&-D(x;L)\partial_{x} \langle I(x;L)\rangle|_{x<x_0}\nonumber\\
  &=&-D(x;0)\partial_{x} \langle I(x;0)\rangle|_{x>x_0}.
\label{eq:S74}
\end{eqnarray}
In addition, because the profile $\langle$$I$$($$x$$;$$0$$)$$\rangle$ is the solution to the generalized diffusion equation with a position-dependent diffusion coefficient $D$$($$x$$;$$0$$)$, the Fick current, $-$$D$$($$x$$;$$0$$)$$\partial_{x}$$\langle$$I$$($$x$$;$$0$$)$$\rangle$, is a constant. A similar result applies to the Fick current, $-$$D$$($$x$$;$$L$$)$$\partial_{x}$$\langle$$I(x;L)$$\rangle$. Thus Eq.~(\ref{eq:S74}) can be further extended to
\begin{eqnarray}
  &&-D(x_1;L)\partial_{x_1} \langle I(x_1;L)\rangle\nonumber\\
  &=&-D(x_2;0)\partial_{x_2} \langle I(x_2;0)\rangle,\quad \forall\, x_{1,2}\in [0,L].
\label{eq:S81}
\end{eqnarray}
Letting $x_1$ be $x$ and $x_2$ be $L-x$, we have
\begin{eqnarray}
  -D(x;L)\partial_{x} \langle I(x;L)\rangle=D(L-x;0)\partial_{x} \langle I(L-x;0)\rangle.\quad\quad
\label{eq:S75}
\end{eqnarray}
On the other hand, Eqs.~(\ref{eq:S28}) and (\ref{eq:S32}) give
\begin{eqnarray}\label{eq:S76}
    D(x;L)=D(L-x;0).
\end{eqnarray}
Because of this Eq.~(\ref{eq:S75}) reduces to
\begin{eqnarray}
  \partial_{x} \left(\langle I(x;L)\rangle+\langle I(L-x;0)\rangle\right)=0
\label{eq:S77}
\end{eqnarray}
and so the quantity $\langle$$I$$($$x$$;$$L$$)$$\rangle$$+$$\langle$$I$$($$L$$-$$x$$;$$0$$)$$\rangle$ is a constant.

To find the value of the constant, we note that since the reflection and transmission for an incident wave on either side of the sample sum to unity,
\begin{eqnarray}
  \langle I(0;x_0)\rangle+\langle I(L;x_0)\rangle=2.
\label{eq:S78}
\end{eqnarray}
On the other hand, from Eq.~(\ref{eq:S73}) the first term is $\langle$$I$$($$0$$;$$L$$)$$\rangle$ and the second is $\langle$$I$$($$L$$;$$0$$)$$\rangle$. Thus
\begin{eqnarray}
  \langle I(0;L)\rangle+\langle I(L;0)\rangle=2,
\label{eq:S79}
\end{eqnarray}
which gives the value of $\langle$$I$$($$x$$;$$L$$)$$\rangle$$+$$\langle$$I$$($$L$$-$$x$$;$$0$$)$$\rangle$ at $x$$=$$0$. Since this value is independent of $x$, we have
\begin{eqnarray}
  \langle I(x;L)\rangle+\langle I(L-x;0)\rangle=2.
\label{eq:S80}
\end{eqnarray}
This is a symmetry relation for the profiles of $\langle$$I$$($$x$$;$$0(L)$$)$$\rangle$. Since the profile of $\langle$$I$$($$x$$;$$x_0$$)$$\rangle$ is constructed in the way described by Eq.~(\ref{eq:S73}), we find that
\begin{equation}\label{eq:S82}
\langle I(x;x_0)\rangle+\langle I(L-x;x_0)\rangle=2
\end{equation}
for ${\rm min}$$(x,L-x)$$<$$x_0$.

\begin{center}
{\bf {\it 2. Explicit expression of $\langle I(x;x_0)\rangle$}}
\end{center}

According to Eq.~(\ref{eq:S73}), in order to find $\langle$$I$$($$x$$;$$x_0$$)$$\rangle$ for arbitrary $x_0$, we only need to find the profiles $\langle$$I$$($$x$$;$$0$$($$L$$)$$)$$\rangle$ for the samples with the reflector placed at the input (output). When the reflector is placed at the input, the generalized diffusion equation (\ref{eq:S104}) reduces to
\begin{eqnarray}
  -\partial_x D(x;0)\partial_x \langle I(x;0)\rangle=0.
\label{eq:S86}
\end{eqnarray}
Its solution can be readily found to be
\begin{eqnarray}
  \langle I(x;0)\rangle&=&\langle I(0^+;0)\rangle-\left(\langle I(0^+;0)\rangle-\langle I(L;0)\rangle\right)\nonumber\\
  &\times&\int_0^x\frac{dx'}{D(x;0)}{\bigg/}\int_0^L\frac{dx'}{D(x;0)}.
\label{eq:S87}
\end{eqnarray}
With the help of Eq.~(\ref{eq:S79}), this reduces to
\begin{eqnarray}
  \langle I(x;0)\rangle&=&\langle I(0^+;0)\rangle-\left(\langle I(0^+;0)\rangle+\langle I(L;0)\rangle-2\right)\nonumber\\
  &\times&\int_0^x\frac{dx'}{D(x;0)}{\bigg/}\int_0^L\frac{dx'}{D(x;0)}.
\label{eq:S88}
\end{eqnarray}
Using Eq.~(\ref{eq:S60}), we find that in $1$D,
\begin{eqnarray}
\label{eq:S89}
\lambda(x;0)=\frac{(x+\zeta)(L-x)}{\ell (L+\zeta)}.
\end{eqnarray}
Combined with the scaling law, this gives
\begin{eqnarray}
\label{eq:S90}
D(x;0)\approx D_0e^{-\frac{(x+\zeta)(L-x)}{\ell (L+\zeta)}}.
\end{eqnarray}
Substituting this into Eq.~(\ref{eq:S88}) and carrying out the integral give
\begin{eqnarray}\label{eq:S91}
    \langle I(x;0)\rangle
    =\langle I(0^+;0)\rangle-\left(\langle I(0^+;0)\rangle+\langle I(0;L)\rangle-2\right)\nonumber \\
   \times\frac{{\rm erf}\left(\sqrt{\frac{sL}{L+\zeta}}\frac{2x-(L-\zeta)}{2L}\right)
   -{\rm erf}\left(-\sqrt{\frac{sL}{L+\zeta}}\frac{L-\zeta}{2L}\right)}
   {{\rm erf}\left(\sqrt{\frac{sL}{L+\zeta}}\frac{L+\zeta}{2L}\right)-{\rm erf}\left(-\sqrt{\frac{sL}{L+\zeta}}\frac{L-\zeta}{2L}\right)},\quad
\end{eqnarray}
where `erf' is the error function and $s$$=$$L$$/$$\ell$. In similar fashion, we can find $\langle$$I$$($$x$$;$$L$$)$$\rangle$, which is
\begin{eqnarray}\label{eq:S92}
    \langle I(x;L)\rangle
    =\langle I(0;L)\rangle-\left(\langle I(0^+;0)\rangle+\langle I(0;L)\rangle-2\right)\nonumber \\
   \times\frac{{\rm erf}\left(\sqrt{\frac{sL}{L+\zeta}}\frac{2x-(L+\zeta)}{2L}\right)
   -{\rm erf}\left(-\sqrt{\frac{sL}{L+\zeta}}\frac{L+\zeta}{2L}\right)}
   {{\rm erf}\left(\sqrt{\frac{sL}{L+\zeta}}\frac{L-\zeta}{2L}\right)-{\rm erf}\left(-\sqrt{\frac{sL}{L+\zeta}}\frac{L+\zeta}{2L}\right)}.\quad
\end{eqnarray}
{Equations (\ref{eq:S91}) and (\ref{eq:S92}) are identical to Eq.~(\ref{eq:12}).}

To determine the value of $\zeta$, we perform numerical experiments for a much larger ensemble with various microscopic parameters matching the real experiment, and then fit {Eq.~(\ref{eq:S73}), where $\langle I(x;0)\rangle$ and $\langle I(x;L)\rangle$ are given by Eqs.~(\ref{eq:S91}) and (\ref{eq:S92}), respectively,} to the results of simulation, with $\zeta$ as the only fitting parameter.

{
\begin{center}
{\bf E. Proof of invariance principle for inhomogeneity arising from nonuniformly distributed scattering strength}
\end{center}
}
We now consider another kind of inhomogeneity. That is, the strength of distributed scattering varies in space. In this case, the sample is characterized by a local mean free path $\ell(x)$, which is the mean free path that would be found in a homogeneously disordered sample with the same disorder structure as at depth $x$. Since the Boltzmann diffusion coefficient $D_0$ is proportional to the mean free path, it varies in space also. It is important to note that, unlike $D(x;x_0)$, this inhomogeneity of the diffusion coefficient has nothing to do with localization effects. When $\xi=\pi\nu D_0$ inherits the spatial inhomogeneity from $D_0$, the action given by Eq.~(\ref{eq:S53}) is modified to
\begin{equation}\label{eq:S102}
\frac{1}{8}\int dx\xi(x){\rm str}(\partial_x Q)^2.
\end{equation}
Correspondingly, Eq.~(\ref{eq:S1}) is modified to
\begin{eqnarray}\label{eq:S105}
    O(x)=\int_{Q(0)=\Lambda}^{Q(L)=Q_L} D[Q]e^{-\frac{1}{8}\int_0^{L} dx\xi(x){\rm str}(\partial_x Q)^2}M_O[Q(x)].\nonumber\\
\end{eqnarray}
Here no tunneling terms appear because the inhomogeneity arises solely from the varying scattering strength.

To proceed, we construct the following map,
\begin{equation}\label{eq:S106}
    x\mapsto \eta(x)=\int_0^x\frac{dx'}{\xi(x')},
\end{equation}
for quasi-$1$D smples, which reduces to
\begin{equation}\label{eq:S113}
    x\mapsto \eta(x)=\int_0^x\frac{dx'}{\ell(x')}
\end{equation}
in $1$D, and substitute it into Eqs.~(\ref{eq:S102}) and (\ref{eq:S105}). This gives
\begin{eqnarray}\label{eq:S107}
    O(\eta)\!\!=\!\!\int_{Q(0)=\Lambda}^{Q(\eta_L)=Q_L}\!\! D[Q]e^{-\frac{1}{8}\int_0^{\eta_L} d\eta{\rm str}(\partial_\eta Q)^2}M_O[Q(\eta)].\quad\quad
\end{eqnarray}
Since this field-theoretical expression is universal with respect to different profiles of $\ell(x)$, this shows that, upon the mapping defined by Eq.~(\ref{eq:S106}) or Eq.~(\ref{eq:S113}), the spatial profile of $O$ in inhomogeneously disordered samples with different spatial variations of $\ell(x)$ must collapse to the same curve.

This universality allows us to calculate local quantities in inhomogeneously disordered materials with a generic profile of $\ell(x)$. For example, to calculate $\langle\ln I(x)\rangle$ in such materials in $1$D, we first perform the mapping of Eq.~(\ref{eq:S113}) to obtain the corresponding uniform disordered sample. For this sample we have \cite{Genack17}
\begin{equation}\label{eq:S109}
    \langle\ln I(\eta)\rangle=-\eta.
\end{equation}
This corresponds to the simulation results shown in the main panel of Fig.~\ref{fig:4}a. Mapping $\eta$ back to $x$, we obtain the profile of $\langle\ln I(x)\rangle$ in the $x$ coordinate, as shown in the inset of Fig.~\ref{fig:4}a. Similarly, we can calculate $\langle I(x)\rangle$. Indeed, for the corresponding uniform disordered sample, we have the generalized diffusion equation,
\begin{equation}\label{eq:S110}
    -\partial_\eta D(\eta)\partial_\eta\langle I(\eta)\rangle=\delta(\eta-\eta'),
\end{equation}
where the position-dependent diffusion coefficient in the $\eta$ coordinate is given by \cite{Tian10}
\begin{equation}\label{eq:S111}
    D(\eta)=e^{-\frac{\eta(\eta_L-\eta)}{\eta_L}}.
\end{equation}
Solving this diffusion equation, we obtain Eq.~(\ref{eq:14}) which we rewrite here,
\begin{eqnarray}\label{eq:S112}
    \langle I(\eta)\rangle
    &=&\langle I(0)\rangle-2\left(\langle I(0)\rangle-1\right)\nonumber \\
   &\times&\frac{{\rm erf}\left(\frac{2\eta-\eta_L}{2\sqrt{\eta_L}}\right)
   -{\rm erf}\left(-\frac{\sqrt{\eta_L}}{2}\right)}
   {{\rm erf}\left(\frac{\sqrt{\eta_L}}{2}\right)-{\rm erf}\left(-\frac{\sqrt{\eta_L}}{2}\right)}.
\end{eqnarray}
This overlaps the simulation results shown in the main panel of Fig.~\ref{fig:4}b. Mapping $\eta$ back to $x$, we obtain the profile of $\langle I(x)\rangle$ in the physical coordinate, as shown in the inset of Fig.~\ref{fig:4}b. In addition, upon the mapping: $x\mapsto\eta(x)$, the probability distribution of intensity for different profiles of $\ell(x)$ must collapse to the same profile, which is confirmed by the simulations shown in Fig.~\ref{fig:4}c.

{
\begin{center}
{\bf F. Proof of invariance principle for general inhomogeneity}
\end{center}
}
In general, inhomogenously disorder structures include multiple ($m$$>$$1$) reflectors and a disordered background with nonuniformly distributed scattering strength. Each reflector labeled by $i$ has transmission coefficient $\Gamma_i$. Let $\{\eta_i\}$ ($i$$=$$1$$,$$2$$,$$\cdots$$,$$m$) be the $\eta$ coordinates of these reflectors. The ensemble average of a local quantity then assumes the general form: $O$$($$\eta$$;$$\{$$\eta_{i}$$\}$$)$. Below we study the variation of $O$ with respect to the changes in $\{$$\eta_i$$\}$, when $\eta$ and $\eta_L$ are fixed.

\begin{center}
{\bf {\it 1. Invariance with regard to the displacement of reflectors with their order unchanged}}
\end{center}

We first consider the case in which the order of reflectors is unchanged when the reflectors are moved. Let the order be $\eta_1$$<$$\eta_2$$<$$\cdots$$<$$\eta_m$ and the number of reflectors in front of $\eta$, denoted $m_l$$($$\eta$$)$, be unchanged. That is, no reflectors cross $\eta$ when moved.

Given the order of the reflectors, $O$ has a field-theoretical expression which is a generalization of Eq.~(\ref{eq:S4}),
\begin{eqnarray}\label{eq:S114}
    &&O(\eta;\{\eta_{i}\})\equiv\int \prod_{i=1}^m DQ_{i-}DQ_{i+} e^{\sum_{i=1}^m\gamma_i {\rm str}(Q_{i-}Q_{i+})}\nonumber\\
    &\times&\prod_{i=0}^m\int_{Q(\eta_i^+)=Q_{i+}}^{Q(\eta_{i+1}^-)=Q_{(i+1)-}} D[Q]e^{-\frac{1}{8}\int_{\eta_i}^{\eta_{i+1}} d\eta{\rm str}(\partial_\eta Q)^2}\nonumber\\
    &\times&M_O[Q(\eta)]
\end{eqnarray}
with $\eta_0$$=$$0$ and $\eta_{m+1}$$=$$\eta_L$, where $Q_{i\pm}$ ($i$$=$$1$$,$$2$$,$$\cdots$$,$$m$) are supermatrices at the two surfaces of the $i$th reflector, $Q_{0+}$$=$$\Lambda$, $Q_{(m+1)-}$$=$$Q_L$, and the coupling constant $\gamma_i$ is obtained from Eq.~(\ref{eq:S97}) with $\Gamma$ replaced by $\Gamma_i$.

For specific $\eta$, we can divide the set $\{$$\eta_{i}$$\}$ uniquely into two disjoint subsets,
\begin{eqnarray}\label{eq:S115}
    \{\eta_{i}\}=\{\eta_{1}<\cdots<\eta_{m_l}\}\cup\{\eta_{m_l+1}<\cdots<\eta_{m}\},\quad
\end{eqnarray}
where the first (second) set are the coordinates of the reflectors in front of (behind) $\eta$. We then rewrite Eq.~(\ref{eq:S114}) in terms of the heat kernel as
\begin{eqnarray}\label{eq:S116}
O(\eta;\{\eta_i\})&=&\int DQ(\eta)\prod_{i=1}^{m} DQ_{{i}+}DQ_{{i}-} e^{-S_t[\{Q{_{{i}\pm}}\}]}\nonumber\\
&\times& A_l
%(\{Q_{{(i\leq m_l)}\pm}\};Q(\eta),\Lambda,\{\eta_{i\leq m_l}\})\quad\quad\quad\nonumber\\
%\times
A_r
%(\{Q_{{(m_l<i\leq m)}\pm}\};Q_L,Q(\eta),\{\eta_{m_l<i\leq m}\}).\quad\quad\quad
M_O[Q(\eta)]
\end{eqnarray}
Here $S_t$ is the generalized tunneling action,
\begin{equation}\label{eq:S117}
    S_t=-\sum_{i=1}^{m}\gamma_{i} {\rm str}(Q_{{i}-}Q_{{i}+}),
\end{equation}
$A_{l}$ is the product of $($$m_l$$+$$1$$)$ heat kernels,
\begin{eqnarray}\label{eq:S118}
A_l&=&W(Q(\eta),Q_{m_l+};\eta-\eta_{m_l})\nonumber\\
&\times&\prod_{i=0}^{m_{l}-1}W(Q_{{(i+1)}-},Q_{{i}+};\eta_{i+1}-\eta_{i}),
\end{eqnarray}
and, $A_{r}$ is the product of $($$m$$-$$m_l$$+$$1$$)$ heat kernels,
\begin{eqnarray}\label{eq:S119}
A_r&=&\prod_{i=m_l+1}^{m}W(Q_{{(i+1)}-},Q_{{i}+};\eta_{i+1}-\eta_{i})\nonumber\\
&\times&W(Q_{{(m_l+1)}-},Q(\eta);\eta_{m_l+1}-\eta).
\end{eqnarray}

Suppose now we displace a reflector and without loss of generality assume that its coordinate is $\eta_{j}$ with $j$$<$$m_l$. Since the order of the reflectors and $m_l$$($$\eta$$)$ are unchanged, the nearest reflectors or the point of observation cannot be crossed. To see how $O$ varies with the change in $\eta_{j}$ we calculate $\partial_{\eta_{j}}$$O$ within the interval, $\eta_{j-1}$$<$$\eta_{j}$$<$$\eta_{j+1}$ (if $j$$=$$m_{l}$ then the upper bound is replaced by $\eta$.). Using Eqs.~(\ref{eq:S116}) and (\ref{eq:S118}) and taking into account Eq.~(\ref{eq:S35}) (with $\gamma$ replaced by $\gamma_{j}$), we obtain
\begin{eqnarray}\label{eq:S120}
\partial_{\eta_{j}}O&=&\int\!\!DQ(\eta)\int\!\!\prod_{i\neq j}^{m} DQ_{{i}+}DQ_{{i}-}e^{-S_t'[\{Q{_{{i}\pm}}\}]}\nonumber\\
&\times&A'_l
%(\{Q_{{i\leq m_l}\pm}\};Q(\eta),\Lambda,\{\eta_{i\leq m_l}\})\nonumber\\
%&\times&
A_r
%(\{Q_{{m_l<i\leq m}\pm}\};Q_L,Q(\eta),\{\eta_{m_l<i\leq m}\})
M_O[Q(\eta)],
\end{eqnarray}
where $S_t'$ excludes the term, $-\gamma_{j}$${\rm str}$$($$Q_{{j}-}$$Q_{{j}+}$$)$, and
\begin{eqnarray}\label{eq:S121}
A'_l=\prod_{i\neq j-1,j}^{m_{l}}W(Q_{{(i+1)}-},Q_{{i}+};\eta_{i+1}-\eta_{i})\quad\quad\quad\nonumber\\
\times\partial_{\eta_{j}}{\tilde W}(Q_{{(j+1)}-},Q_{{(j-1)}+};\eta_{j+1}-\eta_{j-1},\eta_{j}-\eta_{j-1}).\quad\quad
\end{eqnarray}
[We set $Q_{(m_l+1)-}$$=$$Q(\eta)$.] Using Eq.~(\ref{eq:36}) we find that $A'_l$ vanishes for $\eta_{j-1}$$<$$\eta_{j}$$<$$\eta_{j+1}$. Combining this result with Eq.~(\ref{eq:S120}) proves the following identity,
\begin{eqnarray}\label{eq:S122}
    \partial_{\eta_{j}}O(\eta;\{\eta_{i}\})=0,\quad {\rm for}\, \eta_{j-1}<\eta_{j}<\eta_{j+1},
\end{eqnarray}
where the upper bound $\eta_{j+1}$ is replaced by $\eta$ if $\eta_{j}$ is the nearest reflector in front of $\eta$. In the same fashion, we can prove Eq.~(\ref{eq:S122}) for $j$$>$$m_l$.

For the convenience of discussions below, we consider a special case of moving $\eta_j$ ($j$$\leq$$m_l$) to $\eta_{j+1}$ (or $\eta$ for $j$$=$$m_l$). Because of Eq.~(\ref{eq:S122}), the value of $O$ is invariant upon such displacement, which is given by
\begin{eqnarray}\label{eq:128}
O=\int DQ(\eta)\prod_{i\neq j,j\pm 1}^{m} DQ_{{i}+}DQ_{{i}-} e^{-S''_t[\{Q{_{{i}\pm}}\}]}M_O[Q(\eta)]\nonumber\\
\times A_r
%(\{Q_{{(m_l<i\leq m)}\pm}\};Q_L,Q(\eta),\{\eta_{m_l<i\leq m}\})
B\prod_{i\neq j,j\pm 1}^{m_{l}}W(Q_{{(i+1)}-},Q_{{i}+};\eta_{i+1}-\eta_{i}),\quad\quad\quad
\end{eqnarray}
where
\begin{equation}\label{eq:S126}
    S''_t=-\sum_{i\neq j,j+1}^{m}\gamma_{i} {\rm str}(Q_{{i}-}Q_{{i}+}),
\end{equation}
and
\begin{eqnarray}\label{eq:S129}
B&=&\int DQ_{j-}DQ_{(j+1)-}Q_{(j+1)+}\nonumber\\
&\times& W(Q_{j-},Q_{{(j-1)}+};\eta_{j+1}-\eta_{j-1})\nonumber\\
&\times& W(Q_{{(j+2)}-},Q_{(j+1)+};\eta_{j+2}-\eta_{j+1})\nonumber\\
&\times& e^{\gamma_{j} {\rm str}(Q_{{j}-}Q_{(j+1)+})+\gamma_{j+1} {\rm str}(Q_{(j+1)-}Q_{(j+1)-})}.
\end{eqnarray}

\begin{center}
{\bf {\it 2. Invariance with regard to the displacement of reflectors with their order changed}}
\end{center}

We now consider the case in which the order of reflectors in front of or behind the observation point $\eta$ is changed when the reflectors are moved. As before, no reflectors cross $\eta$. Because of Eq.~(\ref{eq:S122}), it is sufficient to consider the case in which the two nearest reflectors in front of or behind $\eta$, say the $j$th and $(j+1)$th reflectors with $j$$\leq$$m_l-1$, are exchanged while all other reflectors are not displaced. The coordinates of the new sequence of reflectors are still denoted $\{\eta_i\}$ with $\eta_1$$<$$\eta_2$$<$$\cdots$$<$$\eta_m$. However, because of the exchange the transmission coefficient of the reflector at $\eta_j$ is $\Gamma_{j+1}$ while at $\eta_{j+1}$ is $\Gamma_{j}$.

Because the order of $\Gamma_{j}$ and $\Gamma_{j+1}$ is exchanged, the field-theoretical expression for the observable, denoted now as $\tilde O$$($$\eta$$;$$\{$$\eta_i$$\}$$)$, is modified to
\begin{eqnarray}\label{eq:S124}
\tilde O(\eta;\{\eta_i\})&=&\int DQ(\eta)\prod_{i=1}^{m} DQ_{{i}+}DQ_{{i}-} e^{-S_t[\{Q{_{{i}\pm}}\}]}\nonumber\\
&\times& \tilde A_l
%(\{Q_{{(i\leq m_l)}\pm}\};Q(\eta),\Lambda,\{\eta_{i\leq m_l}\})\quad\quad\quad\nonumber\\
%\times
A_r
%(\{Q_{{(m_l<i\leq m)}\pm}\};Q_L,Q(\eta),\{\eta_{m_l<i\leq m}\})\quad\quad\quad
M_O[Q(\eta)]
\end{eqnarray}
compared to Eq.~(\ref{eq:S116}), where
\begin{eqnarray}\label{eq:S123}
\tilde A_l&=&\prod_{i\neq j,j\pm 1}^{m_{l}}W(Q_{{(i+1)}-},Q_{{i}+};\eta_{i+1}-\eta_{i})\nonumber\\
&\times&W(Q_{{(j+1)}-},Q_{{(j-1)}+};\eta_{j}-\eta_{j-1})\nonumber\\
&\times&W(Q_{{j}-},Q_{{(j+1)}+};\eta_{j+1}-\eta_{j})\nonumber\\
&\times&W(Q_{{(j+2)}-},Q_{{j}+};\eta_{j+2}-\eta_{j+1}).
\end{eqnarray}
In the same fashion, as the proof of Eq.~(\ref{eq:S122}), we find that Eq.~(\ref{eq:S122}) also holds in the present case. Thus moving $\eta_j$ to $\eta_{j+1}$ renders the value of $\tilde O$ invariant, which is given by
\begin{eqnarray}\label{eq:S125}
\tilde O&=&\int DQ(\eta)\prod_{i\neq j,j\pm 1}^{m} DQ_{{i}+}DQ_{{i}-} e^{-S''_t[\{Q{_{{i}\pm}}\}]}M_O[Q(\eta)]\nonumber\\
&\times& A_r
%(\{Q_{{(m_l<i\leq m)}\pm}\};Q_L,Q(\eta),\{\eta_{m_l<i\leq m}\})
\tilde B\prod_{i\neq j,j\pm 1}^{m_{l}}W(Q_{{(i+1)}-},Q_{{i}+};\eta_{i+1}-\eta_{i}),
\end{eqnarray}
where
\begin{eqnarray}\label{eq:S127}
\tilde B&=&\int DQ_{(j+1)-}DQ_{j-}DQ_{j+}\nonumber\\
&\times& W(Q_{{(j+1)}-},Q_{{(j-1)}+};\eta_{j+1}-\eta_{j-1})\nonumber\\
&\times& W(Q_{{(j+2)}-},Q_{j+};\eta_{j+2}-\eta_{j+1})\nonumber\\
&\times& e^{\gamma_{j} {\rm str}(Q_{{j}-}Q_{{j}+})+\gamma_{j+1} {\rm str}(Q_{(j+1)-}Q_{j-})}.
\end{eqnarray}
By the rotation symmetry, we have $B=\tilde B$.
\begin{equation}\label{eq:S130}
    O(\eta;\{\eta_i\})=\tilde O(\eta;\{\eta_i\}).
\end{equation}
Combining this with Eq.~(\ref{eq:S122}) gives
\begin{eqnarray}\label{eq:S131}
    \partial_{\eta_{j}}O(\eta;\{\eta_{i}\})=0,\quad {\rm for}\, \eta_j\neq\eta.
\end{eqnarray}
On the other hand, the expression Eq.~(\ref{eq:S116}) for $O$ changes when $m_l$ changes. Thus $O$ changes its value when $\eta_j$ is moved across $\eta$. Together with Eq.~(\ref{eq:S131}) this {gives
\begin{eqnarray}\label{eq:S138}
    \forall j:\partial_{\eta_{j}}O(\eta;\{\eta_{i}\})=0,\quad {\rm if\, and\, only\, if}\, \eta_{j}\neq\eta\quad
\end{eqnarray} namely Eq.~(\ref{eq:11}).}

{
\begin{center}
{\bf G. Invariance principle beyond quasi-1D}
\end{center}
}
{The invariance principle is not restricted the quasi-$1$D or $1$D geometries. It applies also to geometries in higher dimension such as a $d$-dimensional slab, for which the sample width is of the order of or much larger than the thickness $L$. To simplify the discussion below, we establish the invariance principle for the inhomogeneity arising solely from an embedded reflector or solely from nonuniformly distributed scattering strength.}

{
\begin{center}
{\bf {\it 1. Inhomogeneity arising from embedded reflector}}
\end{center}
}

{As before, we place a reflector at depth $x_0$ and consider the ensemble average of a generic quantity measured at $(x,y)$, where $x$ is the distance to the input surface and $y$ is the coordinate of the $(d-1)$-dimensional transverse plane. This average quantity, represented by $O(x,y;x_0)$, can be expressed as
\begin{eqnarray}\label{eq:S132}
    &&O(x,y;x_0)\equiv\int D[Q(x_0^-,y)]D[Q(x_0^+,y)]\nonumber\\
    &\times&e^{\gamma_d \int dy{\rm str}(Q(x_0^-,y)Q(x_0^+,y))}\nonumber\\
    &\times&\int_{Q|_{x=0}=\Lambda} D[Q(x<x_0,y)]e^{-\frac{\pi\nu_d D_0}{8}\int dxdy{\rm str}(\nabla Q)^2}\nonumber\\
    &\times&\int_{Q|_{x=L}=Q_L} D[Q(x>x_0,y)]e^{-\frac{\pi\nu_d D_0}{8}\int dxdy{\rm str}(\nabla Q)^2}\nonumber\\
    &\times&M_O[Q(x,y)],
\end{eqnarray}
which is similar to Eq.~(\ref{eq:S4}). Here $\gamma_d=\frac{\omega}{4\pi} \frac{\Gamma}{2+\Gamma}$ and $\nu_d$ is the density of states in $d$ dimension. Comparing this to Eq.~(\ref{eq:S4}), we notice that (i) the supermatrix field $Q(x,y)$ now depends on both $x$ and $y$ except at the input ($x=0$) and output ($x=L$) planes and (ii) the boundary constraints: $Q|_{x=0}=\Lambda$ and $Q|_{x=L}=Q_L$ and the functional $M_O$ stay the same as for the quasi-$1$D case. It is important that Eq.~(\ref{eq:S132}) applies to high-dimensional samples with weak or strong disorder. In the former case, the wave is diffusive while in the latter case, waves are strongly localized ($d=2$) or close to Anderson transition ($d\geq 3$).}

{To demonstrate the invariance principle, we focus below on a $2$D diffusive sample with width equal to $L$ for simplicity. In the present case, the parametrization Eq.~(\ref{eq:S40}) reads
\begin{eqnarray}
\label{eq:S133}
Q(x,y)=e^{w(x,y)}\Lambda e^{-w(x,y)}.
\end{eqnarray}
Expanding the action: $\frac{\pi\nu_d D_0}{8}\int dxdy{\rm str}(\nabla Q)^2$ in $w$, we find that the contributions to the weak localization corrections of $D_0$ from Fourier components: $w_k(x)=\int dy e^{-iky}w(x,y)$ with $k\neq 0$ are much smaller than those from the $k=0$ component. Thus we can reduce Eq.~(\ref{eq:S133}) to $Q(x)=e^{w(x)}\Lambda e^{-w(x)}$. With this substitution we reduce Eq.~(\ref{eq:S132}) to Eq.~(\ref{eq:S4}) and find that $O(x,y;x_0)$ is independent of $y$. Repeating the proof before, we justify Eq.~(\ref{eq:S134}) for $2$D diffusive samples.}

{
\begin{center}
{\bf {\it 2. Inhomogeneity arising from nonuniform distributed scattering strength}}
\end{center}
}

{Let the strength of distributed scattering vary with $x$ and denote the spatially varying mean free path by $\ell(x)$. Since the Boltzmann diffusion coefficient is proportional to the mean free path, it acquires an $x$ dependence denoted $D_0(x)$. Corresponding to this spatial variation, the action given by Eq.~(\ref{eq:S53}) is modified to [cf.~Eq.~(\ref{eq:S102})]
\begin{equation}\label{eq:S135}
\frac{\pi\nu_d}{8}\int dxdy D_0(x){\rm str}(\nabla Q)^2.
\end{equation}
The average quantity can then be expressed as [cf.~Eq.~(\ref{eq:S1})]
\begin{eqnarray}\label{eq:S136}
    O(x,y)&=&\int_{Q|_{{x=0}}=\Lambda}^{Q|_{{x=L}}=Q_L} D[Q]e^{-\frac{\pi\nu_d}{8}\int dxdy D_0(x){\rm str}(\nabla Q)^2}\nonumber\\
    &\times&M_O[Q(x,y)].
\end{eqnarray}
This expression is general and applies to high-dimensional samples with weak or strong disorder.}

{To demonstrate the invariance principle in the present case, we focus on a diffusive sample with the width equal to the thickness $L$. By the same token, when Eq.~(\ref{eq:S133}) is substituted into the action in Eq.~(\ref{eq:S136}), the functional integral is dominated by the field configurations for which $Q$ is uniform in the transverse direction. Thus the action reduces to the action in Eq.~(\ref{eq:S102}), and Eq.~(\ref{eq:S136}) reduces to Eq.~(\ref{eq:S105}). By using the map Eq.~(\ref{eq:S106}), we then establish the invariance principle. That is, the spatial profile of $O$ in inhomogeneously disordered samples with different spatial variations of $\ell(x)$ but the same $\eta_{L}$ collapse to a single curve upon the mapping.}

{The invariance principle established above allows us to control effects of wave coherence in diffusive samples such as the spatial structures of transmission eigechannels, as shown in Fig.~\ref{fig:5}b.}

\begin{center}
{\bf III. Random matrix theory}
\end{center}

{In order to to analytically calculate the complete distribution of the intensity, we apply the methods of RMT. We therefore} in this section give a detailed account of the use of RMT to obtain the statistics of intensity $I(x;x_0)$ in $1$D or in a single mode waveguide,  as in our performed experiments.  As we will show below, the complete
distribution of $I(x;x_0)$ can be obtained within this theoretical model without any fitting parameters. Thus, having the distribution of intensity, we can calculate any average of functions of the intensity. We focus, however, on the statistics of the logarithm of the intensity in a homogeneous sample with a single reflector where some simple analytical results can be given.

\begin{center}
{\bf A. General formalism}
\end{center}

In order to describe multiple scattering processes inside and transmission through a disordered waveguide, we divide the sample into two segments: one in front of and the other behind the observation point $x$ where the intensity is measured. Multiple scattering processes in these two segments are described by the transfer matrices $M_{l}$ and $M_{r}$, respectively, with general structure
\begin{equation}
\label{Eq:S62}
M_{l(r)}=  \begin{bmatrix}
    \alpha_{l(r)} & \beta_{l(r)} \\
      \beta_{l(r)}^*  &  \alpha_{l(r)}^*
\end{bmatrix} ,
\end{equation}
where $\alpha$ and $\beta$ are  complex numbers satisfying $|\alpha|^2-|\beta|^2=1$. In the polar decomposition \cite{Mello-book}, the elements
of the transfer matrices $\alpha_{l (r)}, \beta_{l(r)}$ are written in terms of the phases $\theta_{l (r)}, \mu_{l (r)}$$\in$$[0,2\pi]$ and the radial variables $\lambda_{l(r)}$$\in$$\mathbb{R}^+$. As a result, Eq.~(\ref{Eq:S62}) can written as
\begin{eqnarray}
\label{Eq:S63}
 M_{l (r)} = \begin{bmatrix}
   \sqrt{1+\lambda_{l(r)}}e^{i \theta_{l(r)}}   &  \sqrt{\lambda_{l(r)}}e^{i(2\mu_{l(r)}-\theta_{l(r)})} \\
    \sqrt{\lambda_{l(r)}}e^{-i(2\mu_{l(r)}-\theta_{l(r)})}  & \sqrt{1+\lambda_{l(r)}}e^{-i \theta_{l(r)}}  \nonumber \\
\end{bmatrix}.\nonumber\\
\end{eqnarray}
We model scattering by a reflector embedded inside the random medium by the scattering matrix $S_\Gamma$. Let us assume a
symmetric reflector with transmission coefficient
$\Gamma$. Thus, the reflection coefficient is $1-\Gamma$. Since $S_\Gamma$ is a unitary matrix, it may be written as
\begin{equation}
\label{eq:S64}
S_{\Gamma}=\begin{bmatrix}
    {\sqrt{1- \Gamma}} & i \sqrt{{\Gamma}} \\
     i \sqrt{{\Gamma}}  &  {\sqrt{1-\Gamma}}
\end{bmatrix}.
\end{equation}
This gives the transfer matrix $M_\Gamma$ of the reflector,
\begin{equation}
\label{eq:S65}
M_{\Gamma}= \frac{i}{\sqrt{\Gamma}} \begin{bmatrix}
    1 & - \sqrt{1-\Gamma} \\
      \sqrt{1-\Gamma}  &  -1
\end{bmatrix}.
\end{equation}
For simplicity, throughout this section we assume that the reflector is placed at the input ($x_0$$=$$0$). Equation (\ref{eq:S134}) shows that the result with reflector placed at the input automatically gives the same as for a reflector placed at any point $x_0$ in front of $x$.  A similar expression can be developed in the case of a reflector placed behind the point of observation by assuming that the reflector is at the output surface.

Let $a$ and $b$ be the amplitudes of backwards and forward propagating waves at $x$ within the medium. The intensity is given by
\begin{equation}\label{eq:S61}
    I(x;0)=|a e^{ikx}+be^{-ikx}|^2,
\end{equation}
where $k$ is the wavenumber. $a$ and $b$ can be written  in terms of the elements of $M_{l(r)}$ as follows,
\begin{eqnarray}
\label{eq:S66}
I(x;0)&=&\left({T}/{ T_r} \right) \left|1 -\left(\beta^*_r/\alpha^*_r\right)  e^{-2ikx} \right|^2\nonumber\\
&\equiv& I(M_l, M_r,x),
\end{eqnarray}
where we have used the fact that $T_r$$=$$|$$\alpha_r$$|^{-2}$ is the transmission through the segment behind $x$, while
\begin{equation}
\label{eq:S67}
 T= \left|\alpha_l \alpha^*_r + \beta_l \beta^*_r   \right|^{-2}
\end{equation}
is the transmission through the entire sample. {Equation (\ref{eq:S66}) allows us to make an important observation. Rewriting the equation as
\begin{eqnarray}
\label{eq:S141}
\ln I(x;0)=\ln {T}-\ln { T_r} + \ln \left|1 -\left(\beta^*_r/\alpha^*_r\right)  e^{-2ikx} \right|^2,\quad\quad
\end{eqnarray}we find that there are three contributions to $\ln I$. The first term follows a normal distribution in the deeply localized regime \cite{Beenakker97SM}. But the last two terms are not independent and it cannot be taken for granted that both are negligible or follow a normal distribution. Thus, in general, $\ln I$ does not follow a normal distribution.}

{To find the distribution of $I$, we write Eq.~(\ref{eq:S61}) in terms of the polar variables introduced above as}
\begin{equation}
\label{eq:S68}
 I (x;0) = \frac{I_N(x, \{\lambda,\theta,\mu\})}{I_D(x, \{\lambda,\theta,\mu\})}\equiv I (x, \{\lambda,\theta,\mu\}),
\end{equation}
where
\begin{equation}
 I_N (x, \{\lambda,\theta,\mu\})=\left| \sqrt{1+\lambda_r} -\sqrt{\lambda_r}e^{-2i\left( k x +\mu_r-\theta_r \right)} \right|^2
 \label{eq:S69}
\end{equation}
and
\begin{eqnarray}
 && I_D(x, \{\lambda,\theta,\mu\})=  \frac{1}{\Gamma} \left|   e^{2i(\mu_l+\theta_r)}\sqrt{(1-\Gamma)\lambda_l(1+\lambda_r)} \right. \nonumber  \\
 &&+  \sqrt{(1+\lambda_l)(1+\lambda_r)} e^{2i(\theta_l+\theta_r)} \nonumber  \\
 &&   \left.  + \left( \sqrt{(1-\Gamma)(1+\lambda_l)\lambda_r} +\sqrt{\lambda_l \lambda_r} e^{-2i(\mu_l - \theta_l)}  \right) e^{2i\mu_r}          \right|^2.\nonumber\\
 \label{eq:S70}
\end{eqnarray}

Equation (\ref{eq:S68}) allows us to study the statistical properties of $I(x;x_0)$ by using a scaling
approach to localization and RMT \cite{Mello-book}. Within this framework, the variables: $ \theta_{l(r)}$ and   $\mu_{l(r)}$ in Eq.~(\ref{eq:S68}) are
uniformly distributed, while the distribution of the radial
variable $\lambda_{l(r)}$ is given by \cite{Molina10}
%\begin{eqnarray}
%p_{s_l}(\lambda_l)&=&\frac{C_l}{\left(1+\lambda_l\right)^{1/4}}  \frac{\mathrm{acosh^{1/2}} \sqrt{1+\lambda_l}}{e^{\frac{1}{s_l}\mathrm{acosh}^2 \sqrt{1+\lambda_l}}},\label{poflambda1}\\
% p_{s_r}(\lambda_r)&=&\frac{C_r}{\left(1+\lambda_r\right)^{1/4}}  \frac{\mathrm{acosh^{1/2}} \sqrt{1+\lambda_r}}{e^{\frac{1}{s_r}\mathrm{acosh}^2 \sqrt{1+\lambda_r}}},\label{poflambda}
%\end{eqnarray}
\begin{equation}
\label{eq:poflambda_lr}
p_{s_{l(r)}}\left(\lambda_{l(r)}\right)=C_{l(r)}\frac{\mathrm{acosh^{1/2}}
\sqrt{1+\lambda_{l(r)}}}{\left(1+\lambda_{l(r)}\right)^{1/4}}  e^{-\frac{\mathrm{acosh}^2 \sqrt{1+\lambda_{l(r)}}}{s_{l(r)}}} ,
%%\frac{\mathrm{acosh^{1/2}} \sqrt{1+\lambda_{l(r)}}}{e^{\frac{1}{s_{l(r)}}
%%\mathrm{acosh}^2 \sqrt{1+\lambda_{l(r)}}}},\label{poflambda1}\\
%p_{s_r}(\lambda_r)&=&\frac{C_r}{\left(1+\lambda_r\right)^{1/4}}  \frac{\mathrm{acosh^{1/2}} \sqrt{1+\lambda_r}}{e^{\frac{1}{s_r}\mathrm{acosh}^2 \sqrt{1+\lambda_r}}},\label{poflambda}
\end{equation}
where $C_{l(r)}$ is the normalization constant, $s_l$$=$$x$$/$$\ell$ and $s_r$$=$$($$L-$$x$$)$$/$$\ell$. The closed analytical form for
$p_{s_{l(r)}}\left(\lambda_{l(r)}\right)$ in Eq.~(\ref{eq:poflambda_lr}) is an approximation to the exact
solution of the Melnikov equation, which is given in an integral form \cite{Gertsenshtein59,Genack17}. Equations (\ref{eq:S69}) and (\ref{eq:S70}) are obtained by considering the main contribution to that integral \cite{Melnikov80SM}. The results from Eq.~(\ref{eq:poflambda_lr}) are in practice indistinguishable from those obtained by the exact integral expression \cite{Kleftogiannis13,Chabanov14SM,Rotter14SM}.

\begin{center}
{\bf B. Probability distribution of $\boldsymbol{\ln I(x;x_0)}$}
\end{center}

With the help of Eq.~(\ref{eq:S68}) and the probability distributions of the polar variables introduced above, the distribution of the logarithm of intensity can be written as Eq.~(\ref{eq:13}), which we rewrite here,
\begin{eqnarray}\label{eq:S83}
    P\left(\ln I(x;x_0) \right)=\int\!\!\!\!\int_0^{2\pi}\frac{d\theta_ld\theta_r}{(2\pi)^2}
    \int\!\!\!\!\int_0^{2\pi}\frac{d\mu_ld\mu_r}{(2\pi)^2}\int\!\!\!\!\int_0^\infty d\lambda_ld\lambda_r\nonumber\\
    \times p_{s_l}(\lambda_l)p_{s_r}(\lambda_r)\delta\left(\ln I(x;x_0) - \ln I(x, \{\lambda,\theta,\mu\})\right).\,\qquad
\end{eqnarray}
In Fig.~\ref{fig:2}c of the main text, the distribution of intensity $P\left(\ln I(x;x_0)\right)$ in Eq. (\ref{eq:S83}) is plotted for the reflector placed at
$x_0/L=0$ and 1 for the  observation point at $x/L=1/2$. From the distribution $P\left(\ln I(x;x_0) \right)$, we can obtain other statistical quantities of interest,
although it is difficult to find an explicit analytic expression. In addition, we have calculated the boundary
values $\langle$$I$$($$0$$;$$0$$($$L$$)$$)$$\rangle$ in {Eqs.~(\ref{eq:S91}) and (\ref{eq:S92})} using Eq.~(\ref{eq:S83}). {Note that because of the lack of a closed analytic form of $P\left(\ln I(x;x_0) \right)$ the position-dependent diffusion coefficient $D(x;x_0)$ cannot be obtained from RMT; one must resort to SUSY instead.}

It is of particular interest to study the profile $\langle\ln I(x;x_0)\rangle$, for which we can find an analytic expression. Let the reflector be placed at $x_0$. Taking the logarithm of both sides of Eq.~(\ref{eq:S68}), we obtain
\begin{eqnarray}
\label{eq:S84}
 \ln I (x;x_0) = \ln I_N(x, \{\lambda,\theta,\mu\})-\ln I_D(x, \{\lambda,\theta,\mu\}).
 \nonumber\\
\end{eqnarray}
After averaging the two terms, we obtain from Eq. (\ref{eq:S83})
\begin{eqnarray}
\label{eq:S85}
\langle \ln I(x;x_0) \rangle = \left\{
 \begin{array}{ll}
  -x/\ell  & \textrm{for $x < x_0$  }\\
   -x/\ell +\ln \Gamma & \textrm{for $x > x_0$}
 \end{array}, \right.
\end{eqnarray}
which is Eq.~(\ref{averlnIz}) in the main text.

\end{document}